\documentclass[aps,nofootinbib,preprint]{revtex4}
\usepackage{graphicx}
\usepackage{epsfig}
\usepackage{amsmath}
\usepackage{amssymb}
\newcommand{\slashl}[1]{\not{\!#1}}

\begin{document}

\title{Estimate of the Hadronic Production of the Doubly
Charmed Baryon $\Xi_{cc}$ under GM-VFN Scheme}
\author{Chao-Hsi Chang$^{1,2}$ \footnote{email:
zhangzx@itp.ac.cn}, Cong-Feng Qiao$^{3}$\footnote{email:
qiaocf@gucas.ac.cn}, Jian-Xiong Wang$^{4}$\footnote{email:
jxwang@mail.ihep.ac.cn} and Xing-Gang Wu$^{2}$\footnote{email:
wuxg@itp.ac.cn}}
\address{$^1$CCAST (World Laboratory), P.O.Box 8730, Beijing 100080,
P.R. China.\\
$^2$Institute of Theoretical Physics, Chinese Academy of Sciences,
P.O.Box 2735, Beijing 100080, P.R. China.\\
$^3$Department of Physics, Graduate School of the Chinese Academy
of Sciences, Beijing 100049, P.R. China\\
$^4$Institute of High Energy Physics, P.O.Box 918(4), Beijing
100049, P.R. China}

\begin{abstract}
Hadronic production of the doubly charmed baryon $\Xi_{cc}$
($\Xi^{++}_{cc}$ and $\Xi^{+}_{cc}$) is investigated under the
general-mass variable-flavor-number (GM-VFN) scheme. The gluon-gluon
fusion mechanism and the intrinsic charm mechanisms, i.e. via the
sub-processes $g+g\to(cc)[^3S_1]_{\bar 3}+\bar{c}+\bar{c}$,
$g+g\to(cc)[^1S_0]_6+\bar{c}+\bar{c}$; $g+c\to (cc)[^3S_1]_{\bar
3}+\bar{c}$, $g+c\to (cc)[^1S_0]_6+\bar{c}$ and $c+c\to
(cc)[^3S_1]_{\bar 3}+g$, $c+c\to (cc)[^1S_0]_6+g$, are taken into
account in the investigation, where $(cc)[^3S_1]_{\bar 3}$ (in color
{\bf $\bar 3$}) and $(cc)[^1S_0]_6$ (in color {\bf $6$}) are two
possible $S$-wave configurations of the doubly charmed diquark pair
$(cc)$ inside the baryon $\Xi_{cc}$. Numerical results for the
production at hadornic colliders LHC and TEVATRON show that both the
contributions from the doubly charmed diquark pairs $(cc)[^1S_0]_6$
and $(cc)[^3S_1]_{\bar 3}$ are sizable with the assumption that the
two NRQCD matrix elements are equal, and the total contributions
from the `intrinsic' charm mechanisms are bigger than those of the
gluon-gluon fusion mechanism. For the production in the region of
small transverse-momentum $p_t$, the intrinsic mechanisms are
dominant over the gluon-gluon fusion mechanism and they can raise
the theoretical prediction of the
$\Xi_{cc}$ by almost one order. \\


\end{abstract}
\maketitle

\section{Introduction}

The heavy hadron $\Xi^+_{cc}$ may have been observed by SELEX
Collaboration already\cite{exp,exp2}, although some
comments\cite{comm} pointed out that the measured lifetime is much
shorter and the production rate is much larger than most of the
theoretical predictions
\cite{the,baranov,kiselev1,kiselev2,kiselev3}. It is predicted that
at the fixed target experiment, only about $10^{-5}$ of
$\Lambda_{c}^+$ events in its total sample are producted by
$\Xi^+_{cc}$, however the SELEX collaboration has found that almost
$20\%$ of $\Lambda_{c}^+$ events in its total sample are producted
by $\Xi^+_{cc}$.

In the literature, most of the perturbative QCD (pQCD)
calculations and predictions for $\Xi_{cc}$ \footnote{Throughout
the paper, $\Xi_{cc}$ denotes $\Xi^+_{cc}$ or $\Xi^{++}_{cc}$,
i.e., the isospin-breaking effects are ignorable here.}
hadroproduction are based on the `gluon-gluon fusion mechanism'
i.e. that via the subprocess, $g+g\to(cc)[^3S_1]_{\bar
3}+\bar{c}+\bar{c}$ only. Whereas, the subprocess
$g+g\to(cc)[^1S_0]_6+\bar{c}+\bar{c}$ may also contribute to the
production\cite{majp}. It is because that $\Xi^+_{cc}$ and
$\Xi^{++}_{cc}$ contain the higher components $(ccqg)$ (here
$q=u,d$) etc in their Fock space expansion, so the corresponding
subprocess should be taken into account.

The discussion shown in Ref.\cite{majp} indicates that an
inclusive production rate of $\Xi^{+}_{cc}$ or $\Xi^{++}_{cc}$ can
be factorized into two parts, one part is to produce two free $c$
quarks, which can be calculated by pQCD, another part is to make
these two free $c$ quarks into a $cc$ diquark pair:
$(cc)_{\bf\bar{3}}[^3S_1]$ or $(cc)_{\bf 6}[^1S_0]$, then the
diquark pair hadronizing either into $\Xi^{+}_{cc}$ by absorbing a
quark $d$ or into $\Xi^{++}_{cc}$ by absorbing a quark $u$ for
$(cc)_{\bf\bar{3}}[^3S_1]$, or either into $\Xi^{+}_{cc}$ by
absorbing a quark $d$ or into $\Xi^{++}_{cc}$ by absorbing a quark
$u$ but both absorbing an extra soft gluon for  $(cc)_{\bf
6}[^1S_0]$, all of which can be attributed to non-relativistic QCD
(NRQCD) matrix elements \cite{nrqcd}. In most of the existent
calculations for the hadronic production of $\Xi_{cc}$, the $cc$
diquark pair is assumed to be in $^3S_1$ configuration and in the
color representation ${\bf \bar 3}$ ($(cc)_{\bf\bar{3}}[^3S_1]$).
Whereas according to power counting in velocity $v_c$, the
velocity of the heavy $c$ quarks in the baryon, the NRQCD matrix
elements $h_1$ and $h_3$ (defined in Eq.(\ref{matrixelement})) for
the nonperturbative transition, which correspond to the two
configurations of the diquark pair ($h_1$ is that for
$(cc)_{\bf\bar{3}}[^3S_1]$ and $h_3$ is that for $(cc)_{\bf
6}[^1S_0]$), are at the same order of $v_c$ \cite{majp}. Hence to
give a full estimation of the hadronic production of $\Xi_{cc}$,
we think that $(cc)_{\bf\bar{3}}[^3S_1]$ and $(cc)_{\bf 6}[^1S_0]$
should be treated on the equal footing.

Moreover, as pointed out in Refs.\cite{qiao,zqww}, the so-called
`intrinsic' charm mechanism can give sizable contribution to the
charmonium hadroproduction\cite{qiao}, and to the $B_c$
hadroproduction, especially in small $p_t$ region\cite{zqww}.
Therefore, in addition to considering two configurations of diquark
pair in different color representation ${\bf \bar{3}}$ and ${\bf 6}$
for the gluon-gluon fusion mechanism, it is also interesting to see
how important of the `intrinsic' charm mechanisms via the
sub-processes $g+c\to (cc)[^3S_1]_{\bf \bar 3}+\bar{c}$, $g+c\to
(cc)[^1S_0]_{\bf 6}+\bar{c}$ and $c+c\to (cc)[^3S_1]_{\bf \bar
3}+g$, $c+c\to (cc)[^1S_0]_{\bf 6}+g$, in hadronic production of
$\Xi^+_{cc}$ and $\Xi^{++}_{cc}$ precisely.

Principally, the `intrinsic' charm mechanism induced by the heavy
charm quark is greatly suppressed by the parton distributions in
comparison with the valance and the sea of light quarks and gluon,
but it is `compensated' by `greater phase space' and lower order of
interaction coupling of QCD. Namely the `intrinsic' processes are
$2\to2$ sub-processes at the order of ${\cal O}(\alpha_s^3)$, while
for the gluon-gluon fusion subprocess, its leading contribution
starts at ${\cal O}(\alpha_s^4)$ and is a $2\to 3$ process.

For a fixed target experiment which can reach to the region of very
small transverse momentum $p_t$, the production of the doubly
charmed baryon $\Xi_{cc}$ should additionally involve more
`mechanisms', such as the mechanisms of the so-called intrinsic
charm fusion with the subprocesses $c+c\to (cc)[^3S_1]_{\bf \bar 3}$
and $c+c\to (cc)[^1S_0]_{\bf 6}$, which contribute to the production
only with very small $p_t$ but whose nature essentially is
non-perturbative for QCD. The theoretical predictions on the
hadronic production rate all can be based upon the perturbative QCD
(pQCD) only, though the existent ones are orders of magnitude
smaller than the SELEX observation as pointed out by
Ref.\cite{comm}. Nevertheless, we think that it is worthwhile to
consider more mechanisms than that in the existent predictions, and
to use the updated parton distribution functions (PDFs) in the
general-mass variable-flavor-number (GM-VFN) scheme to re-estimate
the $\Xi_{cc}$ hadroproduction so as to cover a so widen $p_t$
region as pQCD is applicable. Especially, more attention to the
so-called intrinsic charm production mechanism, that is through the
subprocesses $g + c \to (cc)[^3S_1]_{\bf \bar 3}+\bar{c}$, $g+c\to
(cc)[^1S_0]_{\bf 6}+\bar{c}$ and $c+c\to (cc)[^3S_1]_{\bf \bar
3}+g$, $c+c\to (cc)[^1S_0]_{\bf 6}+g$, should be payed.
\footnote{The reliable estimate of the production so far can be that
in terms of pQCD only, so here we take into account all the
mechanisms which are calculable by pQCD. Therefore, here the
so-called intrinsic charm fusion with the subprocesses $c+c\to
(cc)[^3S_1]_{\bf \bar 3}$ and $c+c\to (cc)[^1S_0]_{\bf 6}$ are not
considered (because they are of non-perturbative QCD as mentioned
above). Since the `higher order' mechanisms with the subprocesses:
$c+c\to (cc)[^3S_1]_{\bf \bar 3}+g$, $c+c\to (cc)[^1S_0]_{\bf 6}+g$
are also taken into account so as to `complete the estimate', so in
order to guarantee pQCD applicable and the obtained results being
reliable, we compute the production always to put on a sizable cut
on the transverse momentum $p_t$ of the produced $(cc)$-pair.}

This work is devoted to give a comparative studies of various
production mechanisms, and is also served as a cross-check of the
pQCD calculation for the gluon-gluon fusion mechanism, because the
results given in Ref.\cite{baranov} and Ref.\cite{kiselev1,kiselev2}
are in disagreement. Our results satisfy the gauge invariance at the
amplitude, and our results agree with that of Ref.\cite{baranov}
except for an overall factor 2.

When combining the results of `intrinsic' charm mechanism with the
gluon-gluon fusion mechanism, one needs to make some subtractions
to the `intrinsic' mechanism so as to avoid `double counting'. To
perform the subtraction, we adopt the general-mass
variable-flavor-number (GM-VFN) scheme \cite{acot,gmvfn1,gmvfn2},
in which the heavy-quark mass effects can be treated in a
consistent way both for the hard scattering amplitude and the
PDFs. Moreover, it will be necessary to use the dedicated PDFs
with heavy-mass effects included, which are determined by global
fitting utilizing massive hard-scattering cross-sections. For
instance, for the present analysis, the up-dated one CTEQ6HQ
\cite{6hqcteq} is used.

In Ref.\cite{majp}, the production of $\Xi_{cc}$ at $e^+e^-$
collider is treated carefully and hadronic production is estimated
roughly by comparing with $c$-quark jet both by taking the
fragmentation approach. In the present paper, alternatively, we take
the full pQCD approach to do the estimate with more mechanisms,
because we think the fragmentation approach becomes reliable only at
the high $p_t$ regions where the fragmentation mechanism is dominant
and also the results from the fragmentation approach show a strong
dependence on the input parameter values \cite{majp}. At last, our
results show that when assuming $h_1=h_3$, the contribution to the
Hadronic production of $\Xi_{cc}$ from the doubly charmed diquark
pair in $(cc)_{\bf 6}[^1S_0]$ can be sizable as that from
$(cc)_{\bf\bar{3}}[^3S_1]$.

The paper is organized as follows. In Sec.II, we shall first give
the formulation for the hadronic production of $\Xi_{cc}$ within the
GM-VFN scheme, and then present in some more detail the formulae for
both the gluon-gluon mechanism and the `intrinsic' charm mechanism.
In Sec.III, we present the results for the subprocess and make a
comparison with those in the literature. In Sec.IV, we present the
numerical results for the hadronic production of $\Xi_{cc}$ and make
some discussion over them. The final section is reserved for a
summary.

\section{Formulation under the GM-VFN scheme}

Under the general-mass variable-flavor-number (GM-VFN) scheme
\cite{acot,gmvfn1,gmvfn2}, according to pQCD factorization theorem
the cross-section for the hadronic production of $\Xi_{cc}$ can be
formulated as below:
\begin{eqnarray}
\sigma&=&F^{g}_{H_{1}}(x_{1},\mu) F^{g}_{H_{2}}(x_{2},\mu)
\bigotimes\hat{\sigma}_{gg\rightarrow
\Xi_{cc}}(x_{1},x_{2},\mu)\nonumber\\
&+& \sum_{i,j=1,2;i\neq
j}F^{g}_{H_{i}}(x_{1},\mu)\left[F^{c}_{H_{j}}(x_{2},\mu)-
F^{g}_{H_{j}}(x_{2},\mu)\bigotimes F^c_g(x_2,\mu)\right]
\bigotimes \hat{\sigma}_{gc\rightarrow
\Xi_{cc}}(x_{1},x_{2},\mu)\nonumber\\
&+& \sum_{i,j=1,2;i\neq j}\left[\left(F^{c}_{H_{i}}(x_{1},\mu) -
F^{g}_{H_{i}}(x_{1},\mu)\bigotimes F^{c}_g(x_1,\mu)\right)
\left(F^{c}_{H_{j}}(x_{2},\mu)- F^{g}_{H_{j}}(x_{2},\mu)\bigotimes
F^{c}_g(x_2,\mu)\right)\right]\nonumber\\
&&\bigotimes \hat{\sigma}_{cc\rightarrow
\Xi_{cc}}(x_{1},x_{2},\mu)+\cdots, \label{pqcdf0}
\end{eqnarray}
where the symbol $\cdots$ means even higher order $\alpha_s$
terms. $F^{i}_{H}(x,\mu)$ (with $H=H_1$ or $H_2$; $x=x_1$ or
$x_2$) is the distribution function of parton $i$ in hadron $H$.
$\hat\sigma$ stands for the cross-section of the corresponding
subprocess. For convenience, we have taken the renormalization
scale $\mu_R$ for the subprocess and the factorization scale
$\mu_F$ for factorizing the PDFs and the hard subprocess to be the
same, i.e. $\mu_R=\mu_F=\mu$. In the square bracket, the
subtraction for $F^{c}_{H}(x,\mu)$ is defined as
\begin{equation}\label{subtraction}
F^{c}_{H}(x,\mu)_{SUB}=F^{g}_{H}(x,\mu)\bigotimes
F^{c}_g(x,\mu)=\int^1_{x}F^{c}_g(\kappa,\mu)
F^{g}_{H}\left(\frac{x}{\kappa},\mu\right) \frac{d\kappa}{\kappa}.
\end{equation}
The quark distribution $F^c_g(x,\mu)$ inside an on-shell gluon up
to order $\alpha_s$ can be connected to the familiar $g\to
c\bar{c}$ splitting function $P_{g\to c}$, i.e. $
F^c_g(x,\mu)=\frac{\alpha_s(\mu)}{2\pi}\ln\frac{\mu^2}{m^2_c}P_{g\to
c}(x)$, with $P_{g\to c}(x)=\frac{1}{2}(1-2x+2x^2)$. Later on for
convenience, we shall call the `heavy quark mechanisms', in which
proper subtraction has been given according to method in
GM-VFN scheme, as `intrinsic ones' accordingly.

In Eq.(\ref{pqcdf0}), the first term is the gluon-gluon fusion
mechanism, the second and the third terms are the so called
`intrinsic' charm mechanisms \cite{qiao}, in which all the
subtraction terms are necessary to avoid the double counting
problem, since these terms represent the parts of the gluon-gluon
fusion mechanism which are already included in a fully QCD evolved
`intrinsic' charm distribution function \cite{acot}. The
gluon-gluon fusion mechanism has been considered by several
authors \cite{the,baranov,kiselev1,kiselev2,kiselev3}. However, in
these references, they usually used a PDF in a zero-mass
variable-flavor-number scheme but performed the partonic cross
section calculation using the non-zero heavy-quark masses. Such
treatment shall not heavily affect the results for the hadronic
production at LHC or TEVATRON as is the case of $B_c$ production
\cite{zqww}, however it will make large discrepancies at the fixed
target experiment, i.e. SELEX experiment. This is because, for the
fixed target experiment, most of the generated $\Xi_{cc}$ events
are concentrated in the small $p_t$ regions, where large
uncertainties are caused due to the inconsistent using of PDF.
This is one of the reason that we adopt the GM-VFN scheme to study
the hadronic production of $\Xi_{cc}$ in which the heavy-quark
mass effects can be treated in a consistent way both for the hard
scattering amplitude and the PDFs.

For the `intrinsic' charm mechanisms at the leading order, we need
to calculate subprocesses: $g+c\to \Xi_{cc}^{+}+\bar{c}$ and
$c+\bar{c}\to \Xi_{cc}$. For the hadronic production via
$c+\bar{c}\to \Xi_{cc}$, because its hard subprocess is a $2\to 1$
subprocess and the $p_t$ cut is unavoidable to ensure the
applicable of the PQCD calculation, it at least need to emit one
hard gluon to obtain the $p_t$ distribution. Therefore we shall
calculate $c+\bar{c}\to \Xi_{cc}+g$ other than $c+\bar{c}\to
\Xi_{cc}$ in the following calculations. Note for $c+\bar{c}\to
\Xi_{cc}+g$ mechanism, it has no double counting problem with the
gluon-gluon fusion mechanism and does not need to introduce the
subtraction term, since it is one order higher than the
gluon-gluon fusion mechanism according to the power counting rule
shown in Ref.\cite{acot}.

According to NRQCD formulation, the production rate of $\Xi_{cc}$
can be factorized into two parts, one part is for the production of
two or four free quarks (for the intrinsic mechanism or gluon-gluon
fusion mechanism respectively) and is determined by pQCD, another
part is for non-perturbative transition of the $(cc)$-diquark pair
into $\Xi_{cc}$ and can be defined in terms of non-relativistic QCD
(NRQCD) \cite{nrqcd} matrix elements. According to the discussions
in Ref.\cite{majp}, at the leading order of $v_c$, the baryon
$\Xi_{cc}$ contains two configurations of the $(cc)$-diquark pair,
one is that with the pair in $(cc)_{\bf\bar{3}}[^3S_1]$, another is
that in $(cc)_{\bf 6}[^1S_0]$, whose matrix elements can be written
as
\begin{eqnarray}
h_1&=&\frac{1}{48}\langle 0|[\psi^{a_1} \epsilon \psi^{a_2}+
\psi^{a_2} \epsilon \psi^{a_1}](a^{\dag}a)
\psi^{a_2\dagger} \epsilon \psi^{a_1 \dagger}|0\rangle ,\nonumber\\
h_3&=&\frac{1}{72}\langle 0|[\psi^{a_1} \epsilon \sigma^{i}
\psi^{a_2}- \psi^{a_2} \epsilon \sigma^{i} \psi^{a_1}](a^{\dag}a)
\psi^{a_2\dagger}\sigma^{i} \epsilon \psi^{a_1\dagger}|0\rangle ,
\label{matrixelement}
\end{eqnarray}
where $a_j (j=1,2)$ label the color of the valence quark fields and
$\sigma^{i} (i=1,2,3)$ are Pauli matrices, $\epsilon=i\sigma^2$.
$h_1$ represents the probability for a $(cc)$-diquark pair in
$(cc)_{\bf 6}[^1S_0]$ to transform into the baryon, while $h_3$
represents the probability for a $(cc)$-diquark pair in
$(cc)_{\bf\bar{3}}[^3S_1]$ to transform into the baryon. According
to the discussion in Ref.\cite{majp}, both $h_1$ and $h_3$ are of
order $v_c^2$ to
$|\langle0|\chi^{\dagger}\mathbf{\sigma}\psi|^3S_1\rangle|^2$. The
value of the two matrix elements $h_1$ and $h_3$ can be determined
with non-perturbative methods like QCD sum rule approach, however
their values are unknown yet. The fragmentation of a diquark into a
baryon is assumed to occur with unit probability and consequently,
to have no influence on the production cross section. Further more,
as the fragmentation function $D(z)$ of a heavy diquark into a
baryon, peaks near $z\approx 1$\cite{kiselev1} \footnote{By taking a
simple form of fragmentation function $D(z)$,
Ref.\cite{kiselev2,ref2} did a rough estimation for such effects.
The results there show that such effect is really small.}, and then
the momentum of the final baryon may be considered roughly equal to
the momentum of initial diquark. So to study the hadronic production
of $\Xi_{cc}$ is equivalent to study the hadronic production of
$(cc)$-diquark. Under such condition, the value of NRQCD matrix
element $h_3$ can be naively related to the wave-function for the
color anti-triplet $[^3S_1]$ $cc$ state, i.e.
$h_3=|\Psi_{cc}(0)|^2$. And for convenience, since $h_1$ and $h_3$
is of the same order of $v_c$ \cite{majp}, we take $h_1$ to be $h_3$
hereafter.

\begin{figure}
\centering
\includegraphics[width=0.70\textwidth]{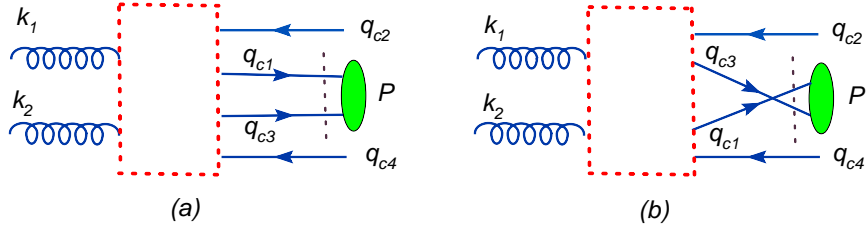}
\caption{The schematic Feynman diagrams for the hadroproduction of
$\Xi_{cc}$ from the gluon-gluon mechanism, where the dashed box
stands for the hard interaction kernel. $k_1$ and $k_2$ are two
momenta for the initial gluons, $q_{c2}$ and $q_{c4}$ are the
momenta for the two outgoing $\bar{c}$, $P$ is the momentum of
$\Xi_{cc}$. The $(cc)$-diquark pair is either in
$(cc)_{\bf\bar{3}}[^3S_1]$ or in $(cc)_{\bf 6}[^1S_0]$
respectively.} \label{feynshort}
\end{figure}

The schematic Feynman diagrams for the gluon-gluon fusion mechanism
are shown in Fig.(\ref{feynshort}). Fig.(\ref{feynshort}) shows that
there are two ways for the two outgoing valence $c$ quarks to form
the $(cc)$-diquark pair and each way contains 36 Feynman diagrams
that are similar to the case of hadronic production of $B_c$ (all
the diagrams can be found in Ref.\cite{bcvegpy1}, and one only need
to change all the $b$ quark line there to the $c$ quark line).
However in Refs.\cite{baranov,kiselev1,kiselev2}, only
Fig.(\ref{feynshort}a) is considered and then only 36 Feynman
diagrams have been taken into consideration.  Since the
contributions from the left and the right diagrams of
Fig.(\ref{feynshort}) are the same and there is an
$\left(\frac{1}{2}\right)$ factor for the square of the amplitude by
taking into account the symmetry of the diquark wavefunction, so
there is an overall factor `2' for our total cross-sections in
comparing with those in Refs.\cite{baranov,kiselev1,kiselev2}. In
the present paper, as a cross check of the results in Refs.
\cite{baranov,kiselev1,kiselev2}, we calculated it by using two
different methods and made a cross check numerically between them.
One method is to fully simplify the amplitude of the gluon-gluon
fusion mechanism by using the improved helicity approach which was
developed in case of the hadronic production of $B_c$
\cite{bcvegpy1,bcvegpy2}. More details of the calculation could be
found in the appendix A. The other is to generate the Fortran
program directly by the Feynman Diagram Calculation (FDC)
program\cite{fdc}, which is a Reduce and Fortran package to perform
Feynman diagram calculation automatically. The detailed treatment
method of $\Xi_{cc}$ in FDC could be found in appendix B.

\begin{figure}
\centering \setlength{\unitlength}{1mm}
\begin{picture}(80,60)(30,30)
\put(-37,-110) {\includegraphics{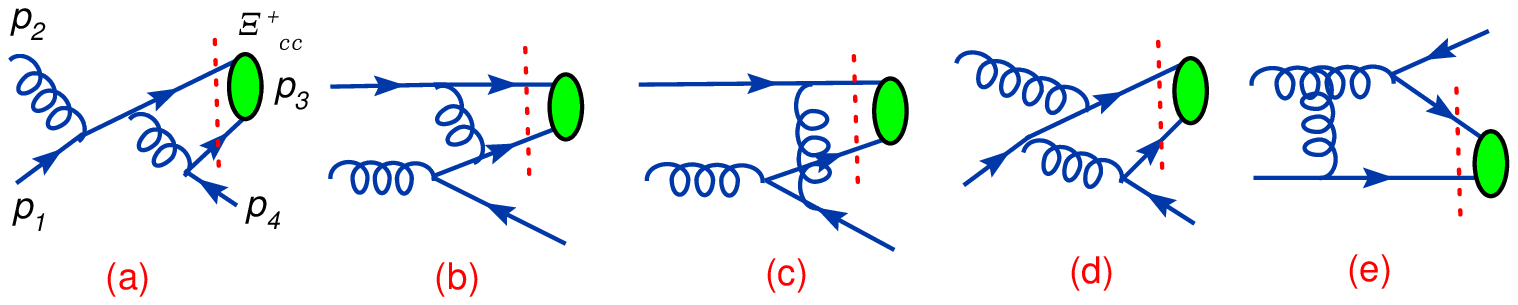}}
\put(-37,-140) {\includegraphics{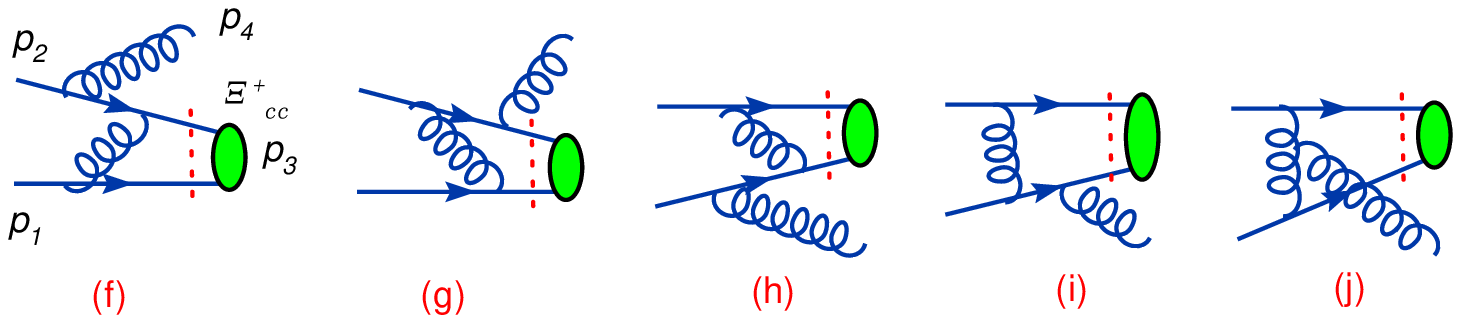}}
\end{picture}
\caption{Typical Feynman diagrams for the sub-processes induced by
`intrinsic' charm. The upper five of them are those for
$g(p_{1})+c(p_{2})\to \Xi_{cc}(p_{3})+\bar{c}(p_{4})$ and the lower
five are for $c(p_{1})+c(p_{2})\to \Xi_{cc}(p_{3})+g(p_{4})$
respectively. The $(cc)$-diquark pair is either in
$(cc)_{\bf\bar{3}}[^3S_1]$ or in $(cc)_{\bf 6}[^1S_0]$
respectively.} \label{fig}
\end{figure}

For the `intrinsic' charm mechanism, we need to consider the
following sub-processes, $g+c\to \Xi_{cc}+\bar{c}$ and
$c+\bar{c}\to \Xi_{cc}+g$, where the $(cc)$-diquark pair in
$\Xi_{cc}$ is in $(cc)_{\bf\bar{3}}[^3S_1]$ or $(cc)_{\bf
6}[^1S_0]$, respectively. Similar to the case of gluon-gluon
fusion mechanism, There is a symmetry factor `2' for cross
section.  The typical Feynman diagrams are shown in
Fig.(\ref{fig}). The final expression of the total square of
amplitude is quit simple, and we adopt the FDC program \cite{fdc}
to obtain it directly.

We will calculate the `intrinsic' charm mechanism within the
GM-VFN scheme. In GM-VFN scheme, when one talks about the heavy
quark components of PDFs, and takes into account of both `heavy
quark mechanisms' and the gluon-gluon fusion mechanism for the
hadronic production, one has to solve the double counting problem:
i.e. a full QCD evolved `heavy quark' charm/bottom distribution
functions, according to the Altarelli-Parisi equations, includes
all the terms proportional to
$\ln\left(\frac{\mu^2}{m^2_Q}\right)$ ($\mu$ the factorization
scale and $m_Q$ the heavy quark mass); and some of them come from
the gluon-gluon fusion mechanism, i.e., a few terms appear from
the integration of the phase-space for the gluon-gluon fusion
mechanism.

To be specific, according to Eq.(\ref{pqcdf0}), the inclusive
$\Xi_{cc}$ hadronic production via `intrinsic charm mechanisms' can
be formulated as,
\begin{equation}
d\sigma=\sum_{ij}\int dx_{1}\int
dx_{2}F^{i}_{H_{1}}(x_{1},\mu_{F})\times
F^{j}_{H_{2}}(x_{2},\mu)d\hat{\sigma}_{ij\rightarrow
\Xi_{cc}X}(x_{1},x_{2},\mu)\,, \label{pqcdf}
\end{equation}
where $i\neq j$ and $i,j=g,c$. Here, the heavy quark PDF
$F^{c}_{H}(x,\mu)$ ($x=x_1$ or $x_2$, $H=H_1$ or $H_2$), should
include a proper subtraction term $F^{c}_{H}(x,\mu)_{SUB}$ as is
defined in Eq.(\ref{subtraction}) in order to avoid the double
counting of $g+c\to\Xi_{cc}+g$ mechanism to the gluon-gluon fusion
mechanism. $d\hat{\sigma}_{ij\rightarrow \Xi_{cc}}$ stands for the
usual 2-to-2 differential cross-section,
\begin{equation}
d\hat{\sigma}_{ij\rightarrow
\Xi_{cc}X}(x_{1},x_{2},\mu^2)=\frac{(2\pi)^4|\overline{M}|^2}
{4\sqrt{(p_1\cdot p_2)^2-m_1^2m_2^2}}
\prod_{i=3}^{4}\frac{d^3\mathbf{p}_i}
{(2\pi)^3(2E_i)}\delta\left(\sum_{i=3}^{4}p_i-p_1-p_2\right),
\end{equation}
where $p_1$, $p_2$ are the corresponding momenta for the initial
two partons and $p_3$, $p_4$ are the momenta for the final ones
respectively. The average over the initial parton's spins and
colors and the sum over the initial and the final state's spins
and colors are absorbed into $|\overline{M}|^2$.
All the expressions of $|\overline{M}|^2$, with all the mass
effects being retained, are shown in the appendix B.

The phase space integration can be manipulated by adopting the routines
RAMBOS~\cite{rkw} and VEGAS~\cite{gpl}, which is the same as that
of the $B_c$ meson generator BCVEGPY~\cite{bcvegpy1,bcvegpy2}.

\section{numerical checks}

Before analyzing the properties for the hadronic production of
$\Xi_{cc}$, we need to check the rightness of program for all the
mechanisms, especially, we should be more careful for the most
complicate gluon-gluon fusion mechanism.

First of all, all the programs are checked by examining the gauge
invariance of the amplitude, i.e. the amplitude vanishes when the
polarization vector of an initial/final gluon is substituted by
the momentum vector of this gluon\footnote{All the Fortran codes
are available from the authors on request.}. Numerically, we find
that the gauge invariance is guaranteed at the computer ability
(double precision) for all these processes.
Next, to make sure the rightness of our program for the
gluon-gluon fusion mechanism, as mentioned before,
the numerical results of our two programs agree with each other exactly.

Furthermore, we compared our numerical results for the gluon-gluon
fusion mechanism with those in the literature by using the same
input parameters as were stated in the corresponding references.  To
make a complete comparison with the results listed in
Ref.\cite{baranov}, we also calculate the partonic cross sections
for the production of $\Xi_{bc}$ and $\Xi_{bb}$ through the
subprocesses, $gg\to\Xi_{bc}+\bar{b}+\bar{c}$ with the
$(bc)$-diquark in color-anti-triplet $[^3S_1]$ or $[^1S_0]$ state,
and $gg\to \Xi_{bb}+\bar{b}+\bar{b}$ with the $(bb)$-diquark in
color-anti-triplet $[^3S_1]$ state. The programs for the production
of $\Xi_{bc}$ and $\Xi_{bb}$ can be easily obtained from the program
for the case of $\Xi_{cc}$.

\begin{figure}
\centering
\includegraphics[width=0.45\textwidth]{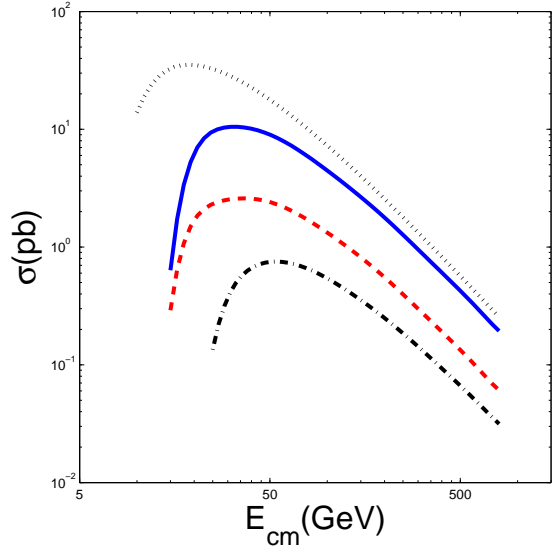}
\caption{The energy dependence of the integrated partonic
cross-section for the production of the baryons with heavy diquarks
via the gluon-gluon fusion mechanism. The dotted line, solid line,
dashed line and dash-dot line stand for the baryons with
$(cc)_{\bf\bar{3}}[^3S_1]$, $(bc)_{\bf\bar{3}}[^3S_1]$,
$(bc)_{\bf\bar{3}}[^1S_0]$ and $(bb)_{\bf\bar{3}}[^3S_1]$
respectively. The curves for $\Xi_{cc}$ and $\Xi_{bb}$ are divided
by 2.} \label{subcs1}
\end{figure}

In Fig.(\ref{subcs1}), we show the partonic cross sections for the
production of baryons with heavy diquarks via the gluon-gluon
fusion subprocess. In drawing the curves, we adopt the same
parameter values as were taken in Ref.\cite{baranov}, i.e.
with a fixed value for $\alpha_s$ ($\alpha_s=0.2$) and
\begin{equation}\label{para1}
|\Psi_{cc}(0)|^2=0.039GeV^3,\;|\Psi_{bc}(0)|^2=0.065GeV^3,\;
|\Psi_{bb}(0)|^2=0.152GeV^3\ ,
\end{equation}
\begin{equation}\label{para2}
m_c=1.8GeV,\; m_b=5.1GeV,\; M_{\Xi_{cc}}=3.6GeV,\;
M_{\Xi_{bc}}=6.9GeV,\; M_{\Xi_{bb}}=10.2GeV \ .
\end{equation}
For convenience of comparison with those of Ref.\cite{baranov}, in
Fig.(\ref{subcs1}), our results for $\Xi_{cc}$ and $\Xi_{bb}$ have
been divided by an overall factor `2'. One may easily find all the
curves for the energy dependence of the partonic cross-sections
shown in Fig.(\ref{subcs1}) are in consistent with the results in
Ref.\cite{baranov} (Fig.(2a) there).

\begin{table}
\caption{Comparison of the partonic cross sections for
$gg\rightarrow \Xi_{cc} +\bar{c}+\bar{c}$ with the corresponding
results in Ref.\cite{kiselev1}, where the $(cc)$-diquark pair is in
$(cc)_{\bf\bar{3}}[^3S_1]$. $E_{cm}$ is the center of mass energy of
the subprocess. The input parameters are $m_c=1.7GeV$,
$M_{\Xi_{cc}}=3.4GeV$, the radial wavefunction at the origin
$R_{cc}(0)=\sqrt{4\pi}\Psi_{cc}(0)=0.601GeV^{3/2}$ and
$\alpha_{s}=0.2$.}
\begin{center}
\begin{tabular}{|c||c|c|c|c|c|c||}
\hline\hline
$E_{cm}$  &15 GeV &20 GeV &40 GeV &60 GeV &80 GeV &100 GeV \\
\hline\hline \(\sigma(pb)\) & $66.6$ & $68.2$ & $41.8$ &
$26.2$ & $17.9$ & $13.1$ \\
\hline \(\sigma(pb)\)\cite{kiselev1} & $23.2$ & $22.5$ &
$13.7$ & $8.96$& $6.45$ &  $4.94$  \\
\hline\hline
\end{tabular}
\label{tabvv}
\end{center}
\end{table}

In Tab.(\ref{tabvv}), we show the comparison of partonic cross
sections (the second column) for the production of $\Xi_{cc}$ with
the $(cc)$-diquark in $(cc)_{\bf\bar{3}}[^3S_1]$ via the gluon-gluon
fusion subprocess with those in Refs.\cite{kiselev1,kiselev2}. In
Tab.(\ref{tabvv}), the results of Ref.\cite{kiselev1} is derived
from the fitted expression (Eq.(8) in Ref.\cite{kiselev1}):
\begin{equation}
\sigma=213.\left(1-\frac{4m_c}{E_{cm}}\right)^{1.9}
\left(\frac{4m_c}{E_{cm}}\right)^{1.35},
\end{equation}
where $E_{cm}$ is the center of mass energy of the subprocess. One
may observe that under the same parameter values, the results in
Refs.\cite{kiselev1,kiselev2} are in disagreement with ours even
though they are close in shape \footnote{Such discrepancy has
already been found in Ref.\cite{baranov}, however the author there
attribute it to the different use of input parameters.}.

\begin{figure}
\centering
\includegraphics[width=0.5\textwidth]{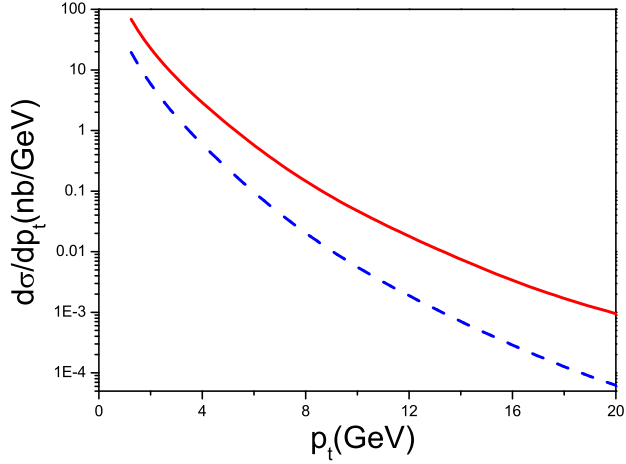}
\caption{The $p_t$-distributions for the `intrinsic' mechanism
$c+c\to\Xi_{cc}+g$ at Tevatron RUN-I $E_{cm}=1.8$TeV (dashed line)
or LHC $E_{cm}=14$TeV (solid line) both with $|y|\leq 1.0$. The
present results are calculated with the same parameters in Ref.
\cite{ref2}.} \label{guter}
\end{figure}

Next, as a cross-check between our results for one of the
`intrinsic' mechanism through $c+c\to\Xi_{cc}+g$ with those of
Ref.\cite{ref2}, we show the cross section of $\Xi_{cc}$-baryon
production at the hadronic energy $E_{cm}=1.8TeV$ or $14TeV$ with
the same input parameters in Fig.(\ref{guter}). The curves in
Fig.(\ref{guter}) agree with those of Ref.\cite{ref2} (Fig.3 and
Fig.4 there).

As a summary, for the gluon-gluon fusion mechanism, except for an
overall factor `2', we confirm the results in Ref.\cite{baranov},
but not those of Ref.\cite{kiselev1,kiselev2}. And for one of the
`intrinsic' charm mechanism, i.e. $c+c\to\Xi_{cc}+g$, our results
agree with those of Ref.\cite{ref2} under the same input
parameters.

\begin{figure}
\centering
\includegraphics[width=0.4\textwidth]{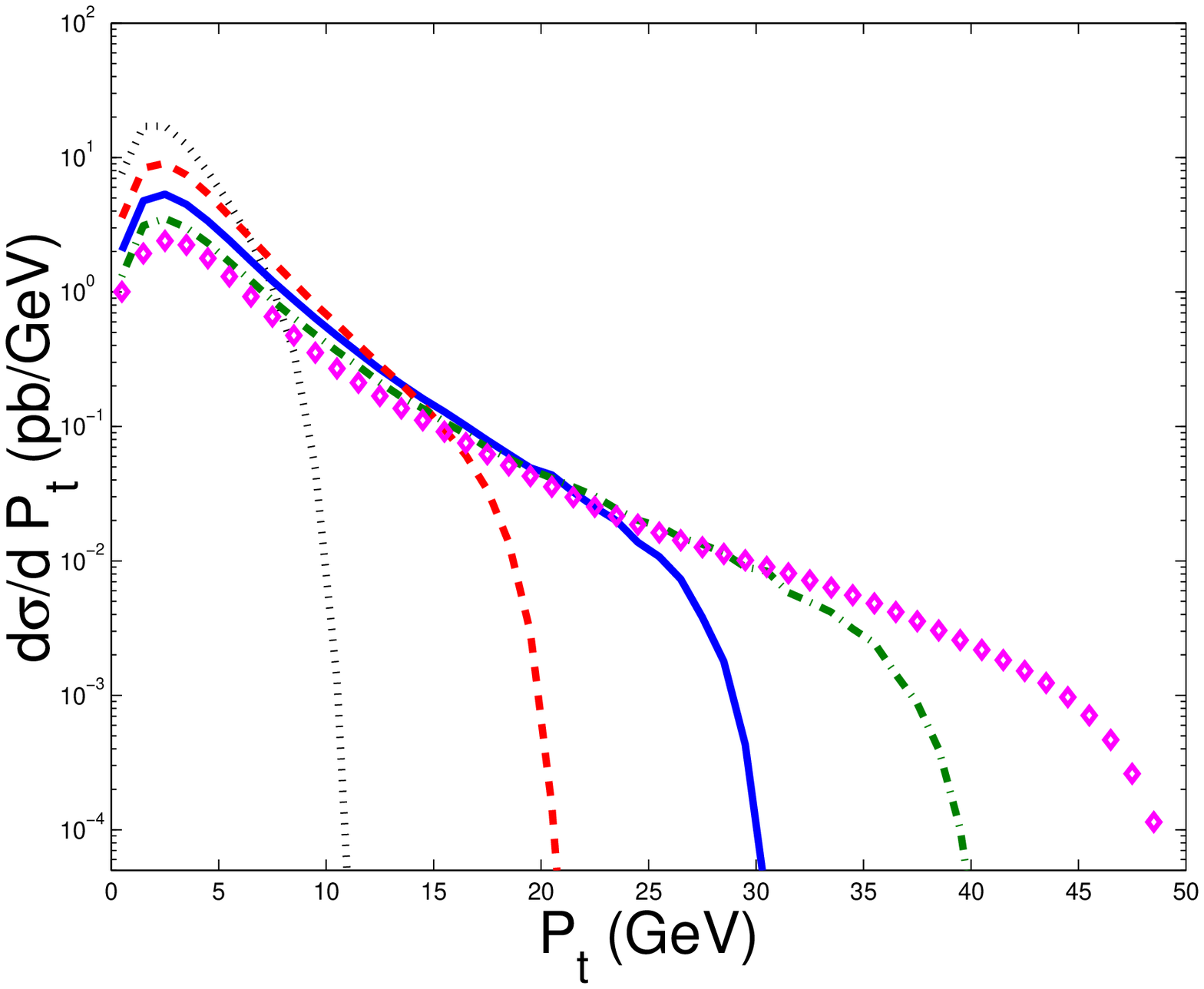}%
\hspace{0.3cm}
\includegraphics[width=0.4\textwidth]{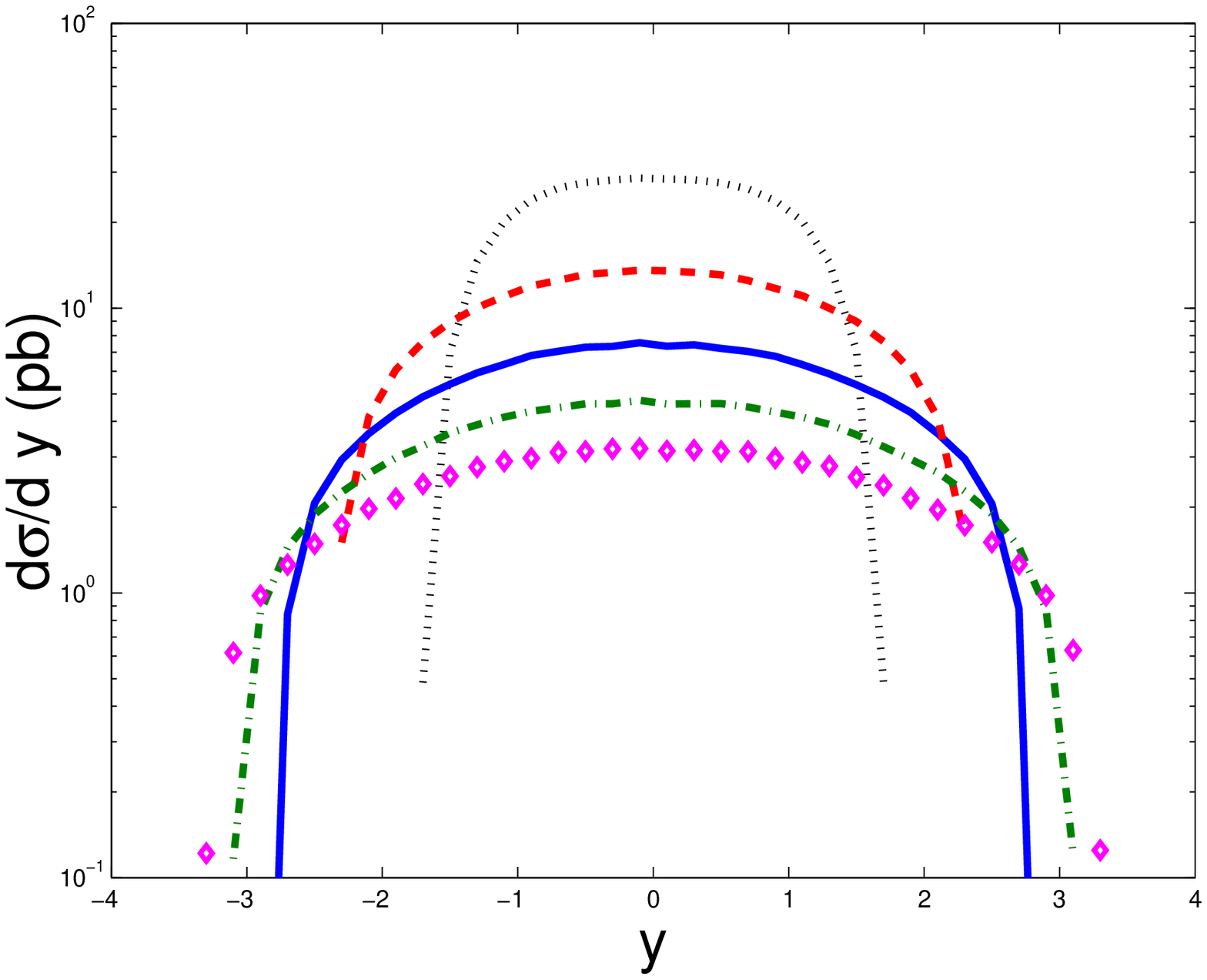}
\caption{The $P_t-$ and $y-$distributions of the produced $\Xi_{cc}$
for the subprocess $gg\to \Xi_{cc}+\bar{c}+\bar{c}$ under different
C.M. energies. Here the $(cc)$ diquark pair only in
$(cc)_{\bf\bar{3}}[^3S_1]$ is taken into account. The dotted line,
dashed line, solid line, dash-dot line and the diamond line stand
for $E_{cm}=20$GeV, $40$GeV, $60$GeV, $80$GeV and $100$GeV,
respectively.} \label{subcs2}
\end{figure}

\begin{figure}
\centering
\includegraphics[width=0.45\textwidth]{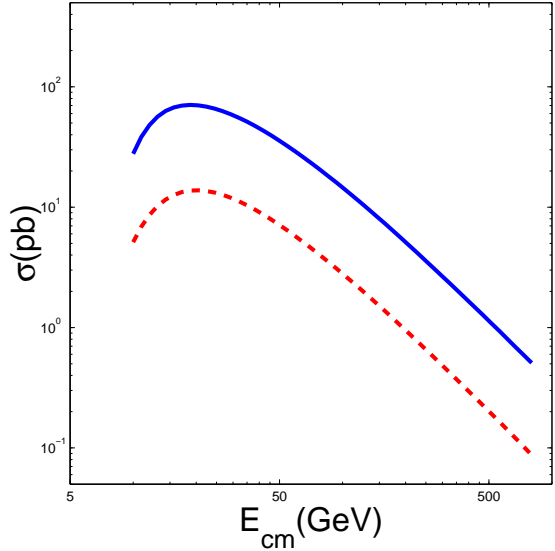}
\caption{The energy dependence of the integrated partonic
cross-section for the $\Xi_{cc}$ production of via the gluon-gluon
fusion mechanism. The solid line and the dashed line stand for the
two $(cc)$-diquark pair states in configurations $(cc)_{\bf\bar
3}[^3S_1]$ and $(cc)_{\bf 6}[^1S_0]$ respectively.} \label{subcs3}
\end{figure}

Finally, we discuss the properties of the two different
configurations of $(cc)$-diquark pair, i.e. $(cc)_{\bf\bar
3}[^3S_1]$ and $(cc)_{\bf 6}[^1S_0]$, for the hadronic production of
$\Xi_{cc}$. In Fig.(\ref{subcs2}), we show the transverse momentum
$P_t$ distribution and the rapidity $y$ distribution at different
center of mass energies for the subprocess
$gg\to\Xi_{cc}+\bar{c}+\bar{c}$, with its $(cc)$-diquark pair in
$(cc)_{\bf\bar{3}}[^3S_1]$. The case of $(cc)$-diquark pair in
$(cc)_{\bf 6}[^1S_0]$ is similar and will not be shown here. We
drawn a comparison between the energy dependence of the integrated
partonic cross-section of the two different $(cc)$-diquark
configurations, i.e. $(cc)_{\bf\bar 3}[^3S_1]$ and $(cc)_{\bf
6}[^1S_0]$, in Fig.(\ref{subcs3}). One may observe that the curves
for $(cc)_{\bf\bar 3}[^3S_1]$ and $(cc)_{\bf 6}[^1S_0]$ are close in
shape and the contributions from $(cc)_{\bf 6}[^1S_0]$ can be up to
$\sim 20\%$ comparing with the case of $(cc)_{\bf\bar 3}[^3S_1]$. So
the contributions from the $(cc)$-diquark pair $(cc)_{\bf 6}[^1S_0]$
are significant and should be taken into consideration for a full
estimation of the hadronic production of $\Xi_{cc}$. This is in
agreement with the conclusion drawn in Ref.\cite{majp}, where the
contributions of these two different states of $(cc)$-diquark pair
are discussed through the fragmentation approach.

\section{hadronic production of $\Xi_{cc}$}

In the present section, we shall first study the hadronic production
properties of $\Xi_{cc}$ both at TEVATRON and at LHC, and then make
a discussion for the hadronic production at the fixed target SELEX
experiment. All the calculations are done under the GM-VFN scheme.

\subsection{hadronic production of $\Xi_{cc}$ at LHC and TEVATRON}

As has been discussed in Sec.II, we take $h_3=|\Psi_{cc}(0)|^2$ and
$h_1=h_3$ in the calculations. The mass of $M_{\Xi_{cc}}$ can be
determined by potential model, and it is estimated to
be\cite{kiselev1}, $M_{\Xi_{cc}}=3.584\pm 0.035 GeV$. In
Ref.\cite{exp}, it has been measured to be $3.519\pm0.001 GeV$. For
clarity, we choose $|\Psi_{cc}(0)|^2=0.039GeV^{3}$\cite{baranov},
$M_{\Xi_{cc}}=3.50 GeV$ and then $m_c=1.75 GeV$. The factorization
energy scale is fixed to be the transverse mass of $\Xi_{cc}$, i.e.
$Q=M_t\equiv \sqrt{M^2+p_{t}^2}$, where $p_t$ is the transverse
momentum of the baryon. The PDFs of version CTEQ6HQ \cite{6hqcteq}
and the leading order $\alpha_s$ running above
$\Lambda^{(n_f=4)}_{QCD}=0.326 GeV$ are adopted.

\begin{table}
\begin{center}
\caption{Cross sections $\sigma$ for the hadronic production of
$\Xi_{cc}$ at TEVATRON and LHC, where the $(cc)$-diquark is in
$(cc)_{\bf\bar 3}[^3S_1]$ or $(cc)_{\bf 6}[^1S_0]$, and the symbol
$g+c$ means $g+c\to \Xi_{cc}+\bar{c}$ and etc. In the calculations,
$p_{t}\geq 4GeV$ is taken and $|y|\leq 1.5$ at LHC, while $|y|\leq
0.6$ at TEVATRON.} \vskip 0.6cm
\begin{tabular}{|c||c|c||c|c||}
\hline - & \multicolumn{2}{|c||}{~~~TEVATRON~($\sqrt S=1.96$
TeV)~~~} & \multicolumn{2}{|c||}{~~~LHC~($\sqrt S=14.0$ TeV)~~~}\\
\hline\hline - & ~~~$(cc)_{\bf\bar 3}[^3S_1]$~~~ & ~~~$(cc)_{\bf
6}[^1S_0]$~~~ & ~~~$(cc)_{\bf\bar 3}[^3S_1]$~~~ & ~~~$(cc)_{\bf
6}[^1S_0]$~~~\\ \hline ~~~$\sigma_{g+g}(nb)$~~~ & ~~~$1.61$~~~ &
~~~$0.392$~~~ & ~~~$22.3$~~~ & ~~~$5.44$~~~ \\ \hline
~~~$\sigma_{c+g}(nb)$~~~ &
~~~$2.29$~~~ & ~~~$0.360$~~~ & ~~~$22.1$~~~ & ~~~$3.42$~~~ \\
\hline ~~~$\sigma_{c+c}(nb)$~~~ & ~~~$0.751$~~~ & ~~~$0.0431$~~~
& ~~~$8.74$~~~ & ~~~$0.475$~~~ \\
\hline\hline
\end{tabular}
\label{totcross}
\end{center}
\end{table}

In TABLE \ref{totcross}, we show the cross-section for the hadronic
production of $\Xi_{cc}$ at TEVATRON and LHC, where $p_{t}\geq 4GeV$
is taken in the calculations and $|y|\leq 1.5$ at LHC, $|y|\leq 0.6$
at TEVATRON. From Tab.\ref{totcross}, one may observe that similar
to the case of hadronic production of $B_c$ meson \cite{zqww}, the
cross-sections of the `intrinsic' charm mechanisms are comparable
to, or even bigger than, the usual considered gluon-gluon fusion
mechanism. From Tab.\ref{totcross}, one may also observe that the
contributions from $(cc)_{\bf 6}[^1S_0]$ are sizable comparing with
that of $(cc)_{\bf\bar 3}[^3S_1]$, i.e. for the gluon-gluon fusion
mechanism, the contribution from $(cc)_{\bf 6}[^1S_0]$ is about
$24\%$ of that of $(cc)_{\bf\bar 3}[^3S_1]$, while for the
mechanisms of $c+g\to\Xi_{cc}+\bar{c}$ and $c+c\to\Xi_{cc}+g$, it
changes to $\sim 15\%$ and $\sim 5\%$, respectively.

\begin{figure}
\centering
\includegraphics[width=0.47\textwidth]{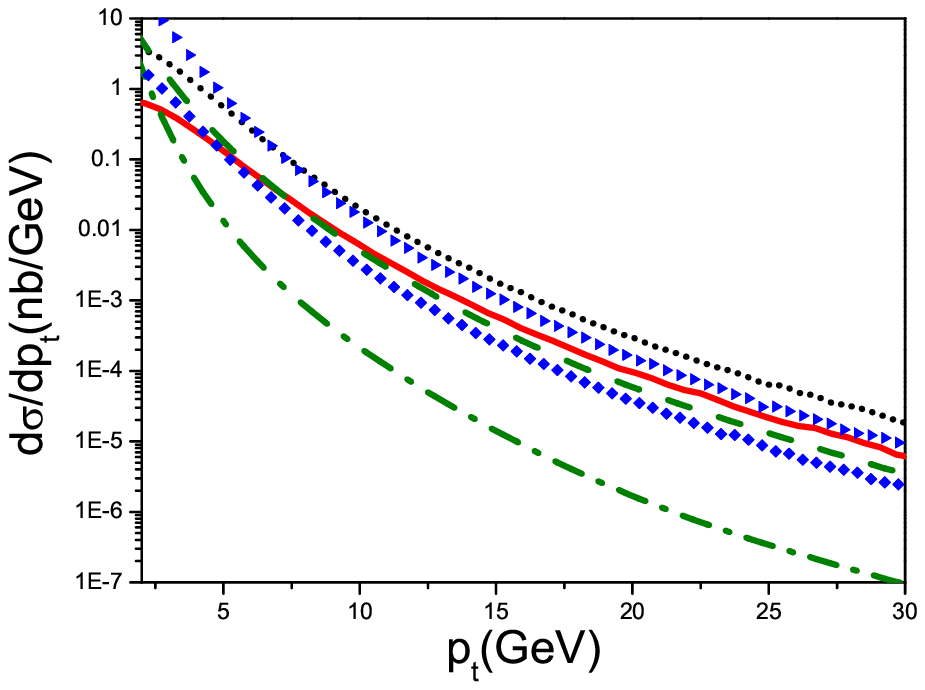}%
\hspace{0.3cm}
\includegraphics[width=0.47\textwidth]{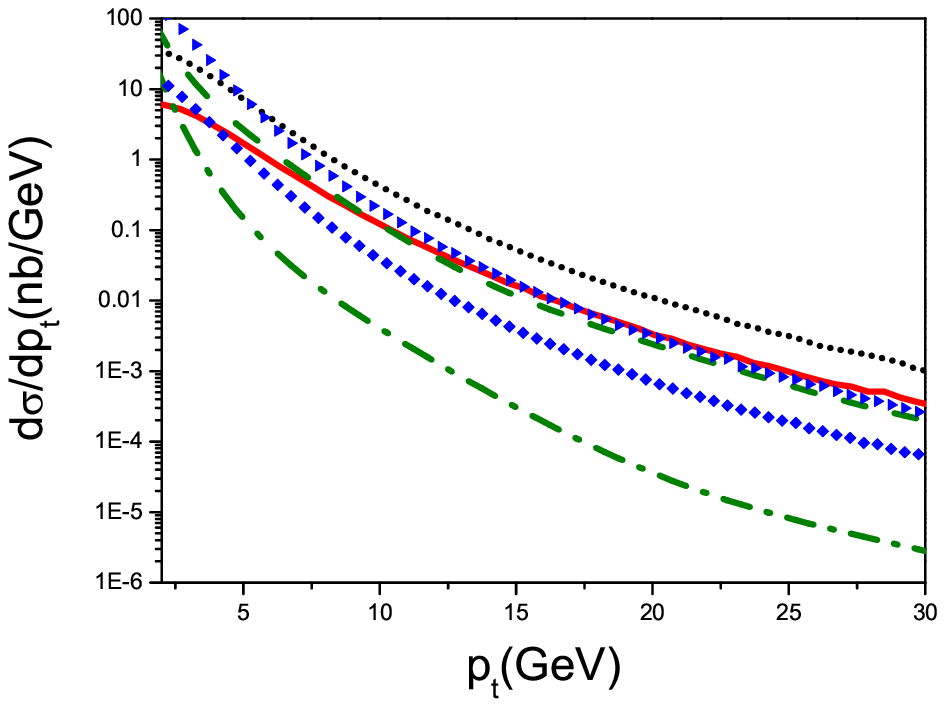}
\caption{The $p_t$-distribution for the hadroproduction of
$\Xi_{cc}$ at TEVATRON (left) and at LHC (right), where $|y|\leq
1.5$ at LHC and $|y|\leq 0.6$ at TEVATRON are adopted. The dotted
line and the solid line are for gluon-gluon fusion mechanism, the
triangle line and the diamond line are for $g+c\to\Xi_{cc}$, the
dashed line and the dash-dot line are for $c+c\to\Xi_{cc}$, where
the upper lines of each mechanism are for $(cc)_{\bf\bar 3}[^3S_1]$
and the lower lines are for $(cc)_{\bf 6}[^1S_0]$, respectively.}
\label{xiccpt}
\end{figure}

In Fig.(\ref{xiccpt}), we show $p_t$ distributions for the hadronic
production of $\Xi_{cc}$ with two configurations of the
$(cc)$-diquark pair states, i.e. $(cc)_{\bf\bar 3}[^3S_1]$ and
$(cc)_{\bf 6}[^1S_0]$, where $|y|\leq 1.5$ at LHC and $|y|\leq 0.6$
at TEVATRON are adopted. From Fig.(\ref{xiccpt}), one may observe
the following points: 1) to compare with the gluon-gluon fusion
mechanism, the `intrinsic' mechanism $g+c\to\Xi_{cc}+g$ dominant in
small $p_t$ regions and its $p_t$ distributions drop faster than
that of gluon-gluon fusion mechanism, which is similar to the case
of $B_c$ hadroproduction \cite{zqww}. 2) For `intrinsic' mechanism
$c+c\to\Xi_{cc}+g$, it $p_t$-distribution drops faster than other
mechanisms and then its contribution is the smallest among all the
mechanisms. 3) For a particular mechanism, the contribution from the
case of $(cc)_{\bf 6}[^1S_0]$ is sizable comparing with the
contribution from the case of $(cc)_{\bf\bar 3}[^3S_1]$. However,
the $p_t$ distribution of $(cc)_{\bf 6}[^1S_0]$ is smaller than that
of $(cc)_{\bf\bar 3}[^3S_1]$ in the whole $p_t$ regions for the same
mechanism and it also drops faster than the case of $(cc)_{\bf\bar
3}[^3S_1]$. Especially for the $c+c\to\Xi_{cc}$ mechanism, $p_t$
distribution of $(cc)_{\bf 6}[^1S_0]$ drop much faster than that of
$(cc)_{\bf\bar 3}[^3S_1]$, and then the cross-section for $(cc)_{\bf
6}[^1S_0]$ is only about $5\%$ of that of
$(cc)_{\bf\bar{3}}[^3S_1]$. As for the gluon-gluon fusion mechanism,
the contribution from $(cc)_{\bf 6}[^1S_0]$ is comparable to that of
$(cc)_{\bf\bar 3}[^3S_1]$ from the `intrinsic' mechanisms at high
energies, especially at LHC, so one should be careful to take the
contribution from $(cc)_{\bf 6}[^1S_0]$ into consideration so as to
provide a full estimation for all these hadronic mechanisms.

\subsection{A simple discussion on hadronic production of $\Xi_{cc}$
at the fixed target SELEX experiment}

For the fixed target experiment, the `intrinsic' charm mechanism
becomes more important than in the case of hadronic production at
TEVATRON or LHC, since small $p_t$ events can contribute here.
Such an experiment has been done by SELEX group \cite{exp} and it
may cover all solid angle without $p_t$ cut, thus the `intrinsic'
charm mechanisms may be studied and extended to very small $p_t$
region. For SELEX experiment, its lower $p_t$ limit can be as
small as $0.2GeV$. However, one should be careful to ensure that
the pQCD calculation is reliable in such small $p_t$ regions, i.e.
the intermediate gluon (with momentum $q$) in all the mechanisms
for the hadronic production of $\Xi_{cc}$ must be hard enough,
i.e. $q^2>>\Lambda_{QCD}^2$.

For the gluon-gluon fusion subprocess, the square of the
intermediate gluon momentum at least is bigger than $(4m_c^2)$ so
as to produce one $c\bar{c}$-quark pair and then is always PQCD
calculable. For the `intrinsic' subprocess $g(p_{1})+c(p_{2})\to
\Xi_{cc}(p_{3})+\bar{c}(p_{4})$, we must ensure that the momentum
of the intermediate gluon of
Fig.(\ref{fig}b,\ref{fig}c,\ref{fig}e) satisfy
\begin{equation}\label{constrainA}
Q^2=-q^2=-\left(p_1-\frac{p_{3}}{2}\right)^2>>\Lambda_{QCD}^2 \;\;
,
\end{equation}
and for the `intrinsic' subprocess $c(p_{1})+c(p_{2})\to
\Xi_{cc}(p_{3})+g(p_{4})$, similarly, we have
\begin{equation}\label{constrainB}
Q^2_1=-q^2_1=-\left(p_1-\frac{p_{3}}{2}\right)^2>>\Lambda_{QCD}^2
\,,\;\;\;\;\;\;
Q^2_2=-q^2_2=-\left(p_2-\frac{p_{3}}{2}\right)^2>>\Lambda_{QCD}^2 .
\end{equation}
Eqs.(\ref{constrainA},\ref{constrainB}) give two extra constraints
for both the partonic fractions $x_1$, $x_2$ and $p_t$. For
definiteness, we set the lowest values for $Q^2$, $Q^2_1$ and
$Q^2_2$ to be $m_c^2$ ($>>\Lambda_{QCD}^2$).

\begin{table}
\begin{center}
\caption{Cross section $\sigma$ for the hadronic production of
$\Xi_{cc}$ at the fixed target experiment with center of mass energy
$33.58GeV$, where the $(cc)$-diquark pair is in $(cc)_{\bf\bar
3}[^3S_1]$ or $(cc)_{\bf 6}[^1S_0]$, and the symbol $g+c$ means
$g+c\to \Xi_{cc}+\bar{c}$ and etc. In the calculations,
$p_{t}>0.2GeV$ is taken.} \vskip 0.6cm
\begin{tabular}{|c||c|c|c||}
\hline - & \multicolumn{3}{|c||}{~~~SELEX ($\sqrt S=33.58$GeV)~~~}\\
\hline\hline - & ~~~$\sigma_{g+g}(pb)$~~~ &
~~~$\sigma_{g+c}(pb)$~~~ & ~~~$\sigma_{c+c}(pb)$~~~ \\ \hline
~~~$(cc)_{\bf\bar 3}[^3S_1]$~~~ & ~~~$4.03$~~~ & ~~~$102.$~~~  & ~~~$1.02\times10^{-3}$~~~ \\
\hline ~~~$(cc)_{\bf 6}[^1S_0]$~~~ & ~~~$0.754$~~~ & ~~~$11.3$~~~ & ~~~$4.15\times10^{-5}$~~~\\
\hline\hline
\end{tabular}
\label{selexcs}
\end{center}
\end{table}

\begin{figure}
\centering
\includegraphics[width=0.50\textwidth]{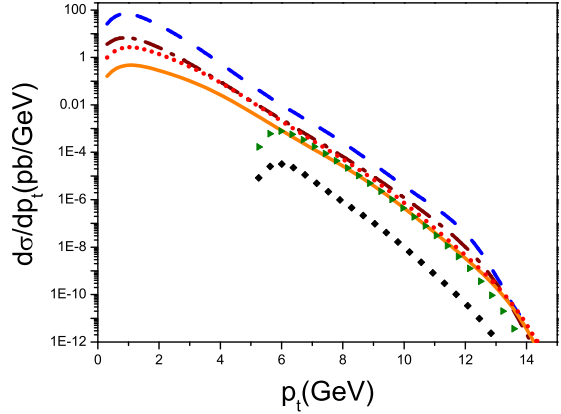}
\caption{The $p_t$-distributions for the hadroproduction of
$\Xi_{cc}$ at SELEX. The dotted line and the solid line are for
gluon-gluon fusion mechanism, the dashed line and the dash-dot line
are for $g+c\to\Xi_{cc}$, the triangle line and the diamond line are
for $c+c\to\Xi_{cc}$, where the upper lines of each mechanism are
for $(cc)_{\bf\bar 3}[^3S_1]$ and the lower lines are for $(cc)_{\bf
6}[^1S_0]$, respectively.} \label{selexpt}
\end{figure}

We show the cross-section for the hadronic production of $\Xi_{cc}$
at SELEX experiment in TABLE \ref{selexcs}, where $p_{t}>0.2GeV$ is
adopted in the calculations. TABLE \ref{selexcs} shows that at
SELEX, the `intrinsic' charm mechanism is the dominant mechanism and
then the theoretical predictions of $\Xi_{cc}$ events at SELEX can
be raised by more than an order. We show the $p_t$ distributions for
the fixed target experiment in Fig.(\ref{selexpt}). One may observe
that the $p_t$ distributions of `intrinsic' mechanism
$g+c\to\Xi_{cc}+\bar{c}$ are bigger than that of the gluon-gluon
fusion mechanism almost in all the $p_t$ region, which is the reason
why the total cross-section of $g+c\to\Xi_{cc}+\bar{c}$ mechanism is
much larger than the gluon-gluon fusion mechanism as shown in TABLE
\ref{selexcs}. For `intrinsic' mechanism $c+c\to\Xi_{cc}+g$, it
$p_t$-distribution starts at $\sim 5GeV$ due to the constraint
Eq.(\ref{constrainB}) and its contribution is quite small. From
TABLE \ref{selexcs}, one may also observe that the contributions
from $(cc)_{\bf 6}[^1S_0]$ are also sizable comparing with that of
$(cc)_{\bf\bar 3}[^3S_1]$ that is similar to the hadronic production
at TEVATRON and LHC as shown in TABLE \ref{totcross}, i.e. for the
gluon-gluon fusion mechanism, the contribution from $(cc)_{\bf
6}[^1S_0]$ is about $19\%$ of that of $(cc)_{\bf\bar 3}[^3S_1]$,
while for the processes of $c+g\to\Xi_{cc}+\bar{c}$ and
$c+c\to\Xi_{cc}+g$, it changes to $\sim 10\%$ and $\sim 4\%$,
respectively.

\section{summary}

We have calculated the hadronic production of the doubly charmed
baryon $\Xi_{cc}$ via the gluon-gluon fusion mechanism and the
`intrinsic' charm mechanism, i.e. via the sub-processes
$g+g\to\Xi_{cc}+\cdots$, $g+c\to \Xi_{cc}+\cdots$ and $c+c\to
\Xi_{cc}+\cdots$. To avoid the double counting problem while taking
the gluon-gluon fusion mechanism and the `intrinsic' charm mechanism
into consideration, we have adopted the GM-VFN scheme in which the
heavy-quark mass effects can be treated in a consistent way both for
the hard scattering amplitude and the PDFs. Some cross checks for
the present results with those in the literature have been done. The
result for the gluon-gluon fusion mechanism agree with what was
given in Ref.\cite{baranov} up to a factor of 2; and the results for
the $c+c\to\Xi_{cc}+g$ with $(cc)$-diquark pair in
$cc_{\bf\bar{3}}[^3S_1]$ agree with that of Ref. \cite{ref2} when
adopting the same input parameters. Whereas the results for the
`intrinsic' mechanisms and those for the cases with $(cc)$-diquark
pair in $cc_{\bf 6}[^1S_0]$ are fresh.

\begin{table}
\begin{center}
\caption{R values, which is defined in Eq.(\ref{Rdef}), for the
hadronic production of $\Xi_{cc}$.} \vskip 0.6cm
\begin{tabular}{|c||c|c|c||}
\hline   & ~~~SELEX~~~ & ~~~TEVATRON~~~ & ~~~LHC~~~ \\
- & ~~~$p_t>0.2GeV$~~~& ~~~$p_t\geq 4GeV$, $|y|\leq 0.6$~~~
& ~~~$p_t\geq 4GeV$, $|y|\leq 1.5$~~~\\
\hline\hline  ~~~$R$~~~ & ~~~$29.$~~~ & ~~~$3.4$~~~ & ~~~$2.8$~~~\\
\hline\hline
\end{tabular}
\label{Rvalue}
\end{center}
\end{table}

From TABLE \ref{totcross} and TABLE \ref{selexcs}, one may see that
the total cross sections of the `intrinsic' charm mechanisms are
comparable to, or even bigger than, that of the gluon-gluon fusion
process, especially for the $g+c\to\Xi_{cc}$ mechanism. To be more
definite, we define a ratio
\begin{equation}\label{Rdef}
R=\frac{\sigma_{total}}{\sigma_{gg\to\Xi_{cc}(
(cc)_{\bf\bar{3}}[^3S_1])}}\  ,
\end{equation}
where $\sigma_{total}$ stands for the cross section for all the
concerned mechanisms and $\sigma_{gg\to\Xi_{cc}
((cc)_{\bf\bar{3}}[^3S_1])}$ is the cross section for the
gluon-gluon fusion mechanism with $(cc)$-diquark pair in
$(cc)_{\bf\bar{3}}[^3S_1]$ configuration only. The values of $R$ for
the hadronic production of $\Xi_{cc}$ in various environments are
shown in Tab.\ref{Rvalue}, which shows that the `intrinsic' charm
mechanisms are not negligible: at SELEX they even dominate over the
other mechanisms. The contributions from the $(cc)$-diquark pair in
$cc_{\bf 6}[^1S_0]$ for all the concerned mechanisms are also
considered in the work, and the results show that if the matrix
element $h_1$ is at the same order of $h_3$ \cite{majp}, i.e.
$h_1\simeq h_3$, the diquark pair will make a sizable contribution
to the hadronic production of $\Xi_{cc}$.

We may conclude that to be a full estimation for the hadronic
production of $\Xi_{cc}$, one needs to take all these mechanisms
into consideration. One may observe that by taking into account the
`intrinsic' mechanisms, the theoretical prediction on the $\Xi_{cc}$
event can be almost one order higher than the previous predictions
in which only the gluon-gluon fusion mechanism is considered.
Nevertheless, there is still a big discrepancy between the SELEX
observation \cite{exp} and pQCD predictions. Perhaps it is due to
the fact that the small $p_t$ regions is not amenable to the pQCD
analysis, e.g., the `intrinsic' mechanism $c+c\to (cc)[^3S_1]_{\bf
\bar 3}+g$ and $c+c\to (cc)[^1S_0]_{\bf 6}+g$, according to
constraint (\ref{constrainB}) there is a big contribution from
non-perturbative QCD range, and the intrinsic charm fusion mechanism
with the subprocesses $c+c\to (cc)[^3S_1]_{\bf \bar 3}$ and $c+c\to
(cc)[^1S_0]_{\bf 6}$, which may contribute to the production greatly
but only with very small $p_t$, and being non-perturbative QCD
nature, it is not considered here. Another possibility might be that
the SELEX experiment does not provide sufficient support for its
claim of evidence for the observation of doubly charmed baryon
$\Xi_{cc}$ as pointed out by Ref.\cite{comm}.

\vspace{10mm}

\noindent {\bf\Large Acknowledgments:} This work was supported in
part by the Natural Science Foundation
of China (NSFC).\\

\appendix

\section{Calculation technology for the gluon-gluon fusion mechanism
under the improved helicity approach}

The general structure of the amplitude in `explicit helicity' form
can be written as
\begin{eqnarray}
M_{i}^{(\lambda_{2}, \lambda_{4}, \lambda_{5},
\lambda_{6})}(q_{c3},q_{c4},q_{c1},q_{c2},k_{1},k_{2})
&=&g_s^4\sum_{\lambda_{2},\lambda_{3}}C_{i}X_{i}D_{1}
B_{Fi}^{(\lambda_{1},\lambda_{2},
\lambda_{3},\lambda_{4},\lambda_{5},\lambda_{6})}
(q_{c3},q_{c4},q_{c1},q_{c2},k_{1},k_{2})
\cdot\nonumber\\
& & D_{2}B_{(cc)}^{(\lambda_{1},\lambda_{3})}(q_{c3},q_{c1}),
\label{matrixd}
\end{eqnarray}
where $i=1,\cdots,72$, $\lambda_{j}\;(j=1,\cdots,6)$ denote the
helicities of the quarks and gluons respectively. $\lambda_{1}$
denotes the helicity of $c(q_{c3})$, $\lambda_{2}$ that of
$\bar{c}(q_{c4})$, $\lambda_{3}$ that of $c(q_{c1})$,
$\lambda_{4}$ that of $\bar{c}(q_{c2})$; whereas $\lambda_{5}$
denotes that of gluon-1 and $\lambda_{6}$ denotes that of gluon-2.
Here $C_i$, $X_i$ denote the color factor and the scalar factor
from all the propagators as a whole for the $i$th-diagram,
respectively. $B_{Fi}^{(\lambda_{1},\lambda_{2},
\lambda_{3},\lambda_{4},\lambda_{5},\lambda_{6})}
(q_{c3},q_{c4},q_{c1},q_{c2},k_{1},k_{2})$ and
$B_{(cc)}^{(\lambda_{1},\lambda_{3})}(q_{c3},q_{c1})$ are the
amplitudes corresponding to the `free quark part'
$g(k_1,\lambda_5)g(k_2,\lambda_6)\rightarrow
c(q_{c3},\lambda_1)+\bar{c}(q_{c4},\lambda_2)+c(q_{c1},\lambda_3)
+\bar{c}(q_{c2},\lambda_4)$ (all the quarks are on-shell) and the
`bound state part' $c(q_{c3},\lambda_1)+c(q_{c1},\lambda_3)
\rightarrow (cc)$, respectively. $D_{1}=\frac{1}{\sqrt{2
q_{c3}\cdot q_{0}}} \frac{1}{\sqrt{2 q_{c4}\cdot q_{0}}}
\frac{1}{\sqrt{2 q_{c1}\cdot q_{0}}} \frac{1}{\sqrt{2 q_{c2}\cdot
q_{0}}}$ and $D_{2}=\frac{1}{\sqrt{2 q_{c1}\cdot q_{0}}}
\frac{1}{\sqrt{2 q_{c3}\cdot q_{0}}}$ are two common normalization
factors.

By comparing Eq.(\ref{matrixd}) with Eq.(22) in
Ref.\cite{bcvegpy1} that is for the $B_c$ hadroproduction, one may
observe that both amplitudes are quite similar with each other.
Most of the present helicity amplitudes can be directly derived
from the results in Ref.\cite{bcvegpy1} by simply replacing the
$b$-quark line there to the present $c$-quark line. And for the
present case, we only need to deal with the following type of the
helicity matrix element (HME) that is quite different from the
case of $B_c$ hadroproduction, i.e.
\begin{equation}
{\rm HME}_i=\langle q_{0\lambda_{2}}|(\slash\!\!\!q_{c4}+
m_{c}){\bf\hat\Gamma_{i}} (\slash\!\!\!q_{c3}-m_{c})|
q_{0\lambda_{1}} \rangle \ ,
\end{equation}
where $i=(1,\cdots,72)$ stands for the $i$-th Feynman diagram and
${\bf \hat\Gamma_{i}}$ means that all the momentum in $\Gamma_{i}$
($\Gamma_{i}$ stands for the explicit strings of Dirac \(\gamma\)
matrices between $\bar{U}(q_{c3})$ and $V(q_{c4})$, which
corresponds to $i$-th Feynman diagram) should change their sign
and the string of the $\gamma$-matrices in $\Gamma_{i}$ should be
written in inverse order. In fact, such type of HME can also be
relate to the familiar one as has been dealt with in the $B_c$
case by adopting the following relation:
\begin{equation}\label{relation}
{\rm HME}_i = -\langle q_{0(-\lambda_{1})}|(\slash\!\!\!q_{c3}+
m_{c})\Gamma_{i} (\slash\!\!\!q_{c4}-m_{c})| q_{0(-\lambda_{2})}
\rangle \ .
\end{equation}
A simple demonstration of Eq.(\ref{relation}) can be found in the
last part of the appendix.

\begin{table}
\begin{center}
\caption{The square of the six independent color factors
(including the cross terms) for
$gg\to(cc)_{\bf\bar{3}}[^3S_1]+\bar{c}+\bar{c}$, ($C_{mij}\times
C_{nij}^*$) with $m,n=(1,2,\cdots,6)$, respectively. }\vspace{3mm}
\begin{tabular}{|c||c|c|c|c|c|c|}
\hline ~ & ~$C_{1ij}^*$~ & ~$C_{2ij}^*$~ & ~$C_{3ij}^*$~ &
~$C_{4ij}^*$~ & ~$C_{5ij}^*$~ & ~$C_{6ij}^*$~ \\
\hline\hline ~$C_{1ij}$~ & $\frac{4}{3}$ & $-\frac{1}{6}$ &
$\frac{2}{3}$ & $-\frac{1}{12}$ & $\frac{5}{12}$ &
$-\frac{1}{3}$  \\
\hline ~$C_{2ij}$~ & $-\frac{1}{6}$ & $\frac{4}{3}$ &
$-\frac{1}{12}$ & $\frac{2}{3}$ & $-\frac{1}{3}$ &
$\frac{5}{12}$\\
\hline ~$C_{3ij}$~ & $\frac{2}{3}$ & $-\frac{1}{12}$&
$\frac{4}{3}$ & $-\frac{5}{12}$& $\frac{1}{12}$ & $-\frac{2}{3}$ \\
\hline ~$C_{4ij}$~ & $-\frac{1}{12}$& $\frac{2}{3}$&
$-\frac{5}{12}$& $\frac{4}{3}$ & $-\frac{2}{3}$ & $\frac{1}{12}$  \\
\hline ~$C_{5ij}$~ & $\frac{5}{12}$& $-\frac{1}{3}$&
$\frac{1}{12}$ & $-\frac{2}{3}$
& $\frac{4}{3}$& $-\frac{1}{6}$  \\
\hline ~$C_{6ij}$~ & $-\frac{1}{3}$& $\frac{5}{12}$&
$-\frac{2}{3}$ & $\frac{1}{12}$
&$-\frac{1}{6}$ & $\frac{4}{3}$\\
\hline
\end{tabular}
\label{colortriplet}
\end{center}
\end{table}

\begin{table}
\begin{center}
\caption{The square of the six independent color factors
(including the cross terms) for $gg\to(cc)_{\bf
6}[^1S_0]+\bar{c}+\bar{c}$, ($C_{mij}\times C_{nij}^*$) with
$m,n=(1,2,\cdots,6)$, respectively.}\vspace{3mm}
\begin{tabular}{|c||c|c|c|c|c|c|}
\hline ~ & ~$C_{1ij}^*$~ & ~$C_{2ij}^*$~ & ~$C_{3ij}^*$~ &
~$C_{4ij}^*$~ & ~$C_{5ij}^*$~ & ~$C_{6ij}^*$~ \\
\hline\hline ~$C_{1ij}$~ & $\frac{8}{3}$ & $-\frac{1}{3}$ &
$\frac{2}{3}$ & $-\frac{1}{12}$ & $\frac{11}{12}$ &
$\frac{1}{6}$  \\
\hline ~$C_{2ij}$~ & $-\frac{1}{3}$ & $\frac{8}{3}$ &
$-\frac{1}{12}$ & $\frac{2}{3}$ & $\frac{1}{6}$ &
$\frac{11}{12}$\\
\hline ~$C_{3ij}$~ & $\frac{2}{3}$ & $-\frac{1}{12}$&
$\frac{8}{3}$ & $\frac{11}{12}$& $-\frac{1}{12}$ & $\frac{2}{3}$ \\
\hline ~$C_{4ij}$~ & $-\frac{1}{12}$& $\frac{2}{3}$&
$\frac{11}{12}$& $\frac{8}{3}$ & $\frac{2}{3}$ & $-\frac{1}{12}$  \\
\hline ~$C_{5ij}$~ & $\frac{11}{12}$& $\frac{1}{6}$&
$-\frac{1}{12}$ & $\frac{2}{3}$
& $\frac{8}{3}$& $-\frac{1}{3}$  \\
\hline ~$C_{6ij}$~ & $\frac{1}{6}$& $\frac{11}{12}$& $\frac{2}{3}$
& $-\frac{1}{12}$ &$-\frac{1}{3}$ & $\frac{8}{3}$\\
\hline
\end{tabular}
\label{colorsixfold}
\end{center}
\end{table}

The sum of all the helicity amplitudes of the sub-process
$g+g\to(cc)+\bar{c}+\bar{c}$ can be arranged as
\begin{eqnarray}
M^{(\lambda_{2}, \lambda_{4}, \lambda_{5},
\lambda_{6})}(q_{c3},q_{c4},q_{c1},q_{c2},k_{1},k_{2})
&=&\sum_{m=1}^{6}C_{mij}M^{(\lambda_{2},\lambda_{4},\lambda_{5},
\lambda_{6})}_{m}(q_{c3},q_{c4},q_{c1},q_{c2},k_{1},k_{2})\;,
\end{eqnarray}
where $C_{mij}$ ($m=1-6$) are six independent color factors of the
process,
\begin{eqnarray}
C_{1ij}&=&\frac{1}{2\sqrt{2}}\left(T^aT^b\right)_{mi}G_{mjk}\;,\;\;
C_{2ij}=\frac{1}{2\sqrt{2}}\left(T^bT^a\right)_{mi}G_{mjk}\;,\nonumber\\
C_{3ij}&=&\frac{1}{2\sqrt{2}}(T^a)_{mj}(T^b)_{ni}G_{mnk}\;,\;\;
C_{4ij}=\frac{1}{2\sqrt{2}}(T^b)_{mj}(T^a)_{ni}G_{mnk}\;,\nonumber\\
C_{5ij}&=& \frac{1}{2\sqrt{2}}\left(T^aT^b\right)_{mj}
G_{mik}\;,\;\; C_{6ij}=\frac{1}{2\sqrt{2}}\left(T^bT^a\right)_{mj}
G_{mik}\;,\label{color}
\end{eqnarray}
where $i,j=1,2,3$ are color indices of the two outgoing anti-quarks
$\bar{c}$ and $\bar{c}$ respectively, and the indices $a$ and $b$
are color indices for gluon-1 and gluon-2 respectively. Here, the
function $G_{mjk}$ equals to the anti-symmetric $\varepsilon_{mjk}$
for the $(cc)$-diquark in $\bf\bar{3}$ configuration and equals to
the symmetric $f_{mjk}$ for the $(cc)$-diquark in $\bf 6$
configuration respectively. The anti-symmetric $\varepsilon_{mjk}$
satisfies $\varepsilon_{mjk}\varepsilon_{m'j'k}=
\delta_{mm'}\delta_{jj'}- \delta_{mj'}\delta_{jm'}$ and the
symmetric $f_{mjk}$ satisfies
$f_{mjk}f_{m'j'k}=\delta_{mm'}\delta_{jj'}+
\delta_{mj'}\delta_{jm'}$.

To get the matrix element squared, one needs to deal with the square
of the above six independent color factors as shown in
Eq.(\ref{color}) (including the cross terms), i.e. ($C_{mij}\times
C_{nij}^*$) with $m,n=(1,2,\cdots 6)$. For reference use, we list
the square of these six independent color factors in TABLE
\ref{colortriplet} and TABLE \ref{colorsixfold}, which are for
$(cc)_{\bf\bar{3}}[^3S_1]$ and $(cc)_{\bf 6}[^1S_0]$, respectively.

By keeping all these points in mind, we rewrite a program based on
the $B_c$ meson generator BCVEGPY\cite{bcvegpy1,bcvegpy2} to
calculate the gluon-gluon fusion mechanism for the hadronic
production of $\Xi_{cc}$.

Finally, we give a simple demonstration of the relation
Eq.(\ref{relation}). To demonstrate the relation
Eq.(\ref{relation}), we shall adopt the following relation,
\begin{equation}\label{helrel}
\langle p_{(\lambda_1)}|\slash\!\!\! k_{1}...\slash\!\!\! k_{n}
|q_{(\lambda_2)}\rangle= (-1)^{n+1}\langle
q_{(-\lambda_2)}|\slash\!\!\! k_{n}...\slash\!\!\! k_{1}
|p_{(-\lambda_1)}\rangle \ ,
\end{equation}
whose non-zero ones can be explicitly written as \cite{zdl}
\begin{eqnarray}
\langle p_{-}|\slash\!\!\! k_{1}...\slash\!\!\! k_{n}
|q_{+}\rangle&=& -\langle q_{-}|\slash\!\!\! k_{n}...\slash\!\!\!
k_{1} |p_{+}\rangle \,\,\,(n \,\,\, even),\\
\langle p_{+}|\slash\!\!\! k_{1}...\slash\!\!\! k_{n}
|q_{-}\rangle&=& -\langle q_{+}|\slash\!\!\! k_{n}...\slash\!\!\!
k_{1} |p_{-}\rangle \,\,\,(n \,\,\, even),\\
\langle p_{+}|\slash\!\!\! k_{1}...\slash\!\!\! k_{n}
|q_{+}\rangle&=& \langle q_{-}|\slash\!\!\! k_{n}...\slash\!\!\!
k_{1} |p_{-}\rangle \,\,\,(n \,\,\, odd),
\end{eqnarray}
where $k_{i}(i=1,\cdots,n)$ are any types of momenta.

Generally, to the $i$-th Feynman diagram, we can expand
$\Gamma_{i}$ as,
\begin{equation}
\Gamma_{i}=\sum_n C_n(\slashl{p}_1\slashl{p}_2\cdots\slashl{p}_n),
\end{equation}
and then we have,
\begin{equation}
{\bf \hat\Gamma_{i}}=\sum_n (-1)^n C_n(\slashl{p}_n\cdots
\slashl{p}_{2}\slashl{p}_1),
\end{equation}
where $C_n$ are functions free of Dirac $\gamma$ matrix element.
Taking use of Eq.(\ref{helrel}), we finally obtain
\begin{eqnarray}
&& \langle q_{0(-\lambda_{1})}|(\slash\!\!\!q_{c3}+
m_{c})\Gamma_{2i}
(\slash\!\!\!q_{c4}-m_{c})| q_{0(-\lambda_{2})} \rangle \nonumber\\
&=&\sum_n C_n \langle q_{0(-\lambda_{1})}|
\left[\slashl{q}_{c3}(\slashl{p}_1\slashl{p}_2\cdots\slashl{p}_n)
\slashl{q}_{c4} -m_c\slashl{q}_{c3}(\slashl{p}_1\slashl{p}_2
\cdots\slashl{p}_n)\right.\nonumber\\
&& \quad\quad\quad\quad\quad\quad\quad + \left.
m_c(\slashl{p}_1\slashl{p}_2\cdots
\slashl{p}_n)\slashl{q}_{c4}-m_c^2(\slashl{p}_1
\slashl{p}_2\cdots\slashl{p}_n) \right] | q_{0(-\lambda_{2})}
\rangle\nonumber\\
&=&\sum_n C_n \langle q_{0(\lambda_{2})}|
\left[(-1)^{n+3}\slashl{q}_{c4}(\slashl{p}_n\cdots\slashl{p}_2
\slashl{p}_1) \slashl{q}_{c3}
-(-1)^{n+2}m_c(\slashl{p}_n\cdots\slashl{p}_2
\slashl{p}_1)\slashl{q}_{c3}\right.\nonumber\\
&& \quad\quad\quad\quad\quad + \left.
(-1)^{n+2}m_c\slashl{q}_{c4}(\slashl{p}_n\cdots\slashl{p}_2
\slashl{p}_1)-(-1)^{n+1}m_c^2(\slashl{p}_n\cdots\slashl{p}_2
\slashl{p}_1) \right] | q_{0(\lambda_{1})} \rangle\nonumber\\
&=&-\langle q_{0(\lambda_{2})}|(\slash\!\!\!q_{c4}+
m_{c}){\bf\hat\Gamma_{2i}} (\slash\!\!\!q_{c3}-m_{c})|
q_{0(\lambda_{1})} \rangle .
\end{eqnarray}

\section{calculation technology in FDC program\cite{fdc} and
the square of amplitude for the intrinsic charm mechanism}

First, we take gluon-gluon fusion mechanism as an explicit example
to show the technology used in FDC program\cite{fdc} and show in
more detail how we can derive the program for the hadronic
production of $\Xi_{cc}$ from those of $J/\psi$.

The amplitude for each Feynman diagram of $g+g\to J/\psi(p_3) +
c(p_4) + {\bar c}(p_5)$ can be written as:
\begin{eqnarray}
&M(J/\psi)=& {\bar u(p_4,s_4)}\Gamma_1 s_f(k_1,m_c) \cdots
s_f(k_{n-1},m_c) \Gamma_n v(\displaystyle{p_3 \over
2},s_1)\nonumber\\
& &B(p_3,s,s_1,s_2,m_{J/\psi}) {\bar u(\displaystyle{p_3 \over
2},s_2)} \Gamma^\prime_1 s_f(q_1,m_c) \cdots s_f(q_{n^\prime -1})
\Gamma^\prime_{n^\prime} v(p_5,s_5).\label{jpsiline}
\end{eqnarray}
where $s_f(k,m)$ ($k=k_i$ or $q_i$) is the fermion propagator,
$B(p_3,s,s_1,s_2,m_{J/\psi})$ is the wavefunction of $J/\psi$,
$\Gamma_1,\cdots,\Gamma_n, \Gamma^\prime_1,
\cdots,\Gamma^\prime_n,$ are the interaction vertices. The color
factor part is treated separately (similar to the method described
in Appendix.A) and will not discussed here.

One can easily find out the corresponding Feynman diagram in $g+g\to
\Xi_{cc}(p_3) + {\bar c}(p_4) + {\bar c}(p_5)$ and the amplitude of
it could be written as:
\begin{eqnarray}
&M(\Xi_{cc})=&{\bar u(\displaystyle{p_3 \over 2},s_1)} \Gamma_n
s_f(-k_{n-1},m_c) \cdots s_f(-k_1,m_c) \Gamma_1 v(p_4,s_4)
\nonumber\\
& &B(p3,s,s_1,s_2,m_{\Xi_{cc}}) {\bar u(\displaystyle{p_3 \over
2},s_2)} \Gamma^\prime_1 s_f(q_1,m_c) \cdots s_f(q_{n^\prime
-1},m_c) \Gamma^\prime_{n^\prime} v(p_5,s_5),\label{xiccline}
\end{eqnarray}
where $B(p_3,s,s_1,s_2,m_{\Xi_{cc}})$ is the wavefunction of
$\Xi_{cc}$. For an arbitrary Fermion line,
\begin{displaymath}
a={\bar u(\displaystyle{p_3 \over 2},s_1)} \Gamma_n
s_f(-k_{n-1},m_c) \cdots s_f(-k_1,m_c) \Gamma_1 v(p_4,s_4),
\end{displaymath}
we have
\begin{eqnarray*}
&&a=a^T=v^T(p_4,s_4) \Gamma^T_1 s^T_f(-k_1,m_c) \cdots
s^T_f(-k_{n-1},m_c) \Gamma^T_n
      {\bar u(\displaystyle{p_3 \over 2},s_1)}^T \\
&&=v^T(p_4,s_4) C C^- \Gamma^T_1 C C^- s^T_f(-k_1,m_c) C C^- \cdots
   C C^- s^T_f(-k_{n-1},m_c) C C^-\Gamma^T_n CC^- {\bar u(\displaystyle{p_3 \over 2},s_1)}^T \\
&&=(-1)^{(n+1)} {\bar u(p_4,s_4)}\Gamma_1 s_f(k_1,m_c) \cdots
s_f(k_{n-1},m_c) \Gamma_n v(\displaystyle{p_3 \over 2},s_1),
\end{eqnarray*}
with the help of the following equations
\begin{eqnarray*}
&v^T(p_4,s_4) C = -{\bar u(p_4,s_4)},~~
&C^- {\bar u({p_3 \over 2},s_1)}^T = v({p_3 \over 2},s_1),~~ \\
&C^- \Gamma^T_i C = - \Gamma_i,~~ &C^- s^T_f(-k_i,m_c) C =
s_f(k_i,m_c) .
\end{eqnarray*}
Where $C=-i\gamma^2\gamma^0$ is the charge conjugation matrix.
And then Eq.(\ref{xiccline}) can be transformed as
\begin{eqnarray}
&M(\Xi_{cc})=&(-1)^{(n+1)}{\bar u(p_4,s_4)}\Gamma_1 s_f(k_1,m_c)
\cdots s_f(k_{n-1},m_c) \Gamma_n v(\displaystyle{p_3 \over 2},s_1)
\nonumber\\
&& B(p_3,s,s_1,s_2,m_{\Xi_{cc}}) {\bar u(\displaystyle{p_3 \over
2},s_2)} \Gamma^\prime_1 s_f(q_1,m_c) \cdots s_f(q_{n^\prime -1})
\Gamma^\prime_{n^\prime} v(p_5,s_5).\label{newline}
\end{eqnarray}

By comparing Eq.(\ref{jpsiline}) with Eq.(\ref{newline}), one find
that they are the same except for an overall factor
$(-1)^{(n+1)}$, where `$n$' is the interaction vertex number of
the corresponding fermion line and depends on the detail of each
Feynman diagram. Therefore, we can completely use the method of
$J/\psi$ to deal with $\Xi_{cc}$ case by adding a factor
$(-1)^{(n+1)}$ diagram by diagram. The detailed description of
method to treat the $J/\psi$ and $B_c$ calculation could be found
in the Ref.\cite{fdc,bcvegpy2}.

All the above discussion is also valid for the calculation of the
`intrinsic' charm mechanisms. And the following results are
obtained by taking the FDC program.

For convenience, we express the square of the amplitudes by the
variants $s$, $t$ and $u$, which are defined as:
\begin{displaymath}
s=(p_{1}+p_{2})^2,~~~t=(p_{1}-p_{3})^2,~~~u=(p_{1}-p_{4})^2 ,
\end{displaymath}
where $p_i=(E_i,p_{ix},p_{iy},p_{iz})$ are the corresponding momenta
for the involved particles: $p_1$ and $p_2$ are the momenta of
initial partons, $p_3$ and $p_4$ are the momenta of $\Xi_{cc}$ and
another outgoing particles respectively. Further more, for
$\bar{c}(p_1)+g(p_2)\to \Xi_{cc}(p_3)+\bar{c}(p_4)$, we set
\begin{displaymath}
u_{1}=(u-4m_c^2),~~~s_{1}=(s-m_{c}^2),~~~t_{1}= (t-m_{c}^2) ,
\end{displaymath}
and for $c(p_1)+c(p_2)\to \Xi_{cc}(p_3) +g(p_4)$, we set
\begin{displaymath}
u_{1}=(u-m_{c}^2),~~~s_{1}=(s-4m_c^2),~~~t_{1}= (t-m_{c}^2).
\end{displaymath}
The relation, $u_1+t_1+s_1=0$, is useful to make all the
expressions for the square of the amplitudes compact.

The square of the amplitude for the subprocess $c(p_1)+g(p_2)\to
\Xi_{cc}(p_3)+\bar{c}(p_4)$ with $(cc)$-diquark pair in
$(cc)_{\bf\bar 3}[^3S_1]$ can be written as,
\begin{eqnarray}
|\overline M|^2 &=&{2^9{\alpha_s}^3|\Psi_{cc}(0)|^2{\pi}^4 \over
3^5{M}}\left[ 4M^2\left({10 \over u_{1}^2}+{-4 \over
s_{1}t_{1}}+{11u_{1}^2 \over s_{1}^2t_{1}^2}+{4 u_{1}^4 \over
s_{1}^3t_{1}^3}\right)+4M^4 \left({-17 \over s_{1}
t_{1}u_{1}}+{28u_{1} \over
s_{1}^2t_{1}^2}+{-20 u_{1}^3 \over s_{1}^3t_{1}^3}\right)\right.   \nonumber\\
& & \left. +3M^6\left({-12 \over s_{1}t_{1}u_{1}^2}+{-5 \over
s_{1}^2t_{1}^2}+ {-14u_{1}^2 \over s_{1}^3t_{1}^3}+{4u_{1}^4 \over
s_{1}^4t_{1}^4}\right)+8\left({-2 \over u_{1}}+{11u_{1} \over
s_{1}t_{1}}+{-9u_{1}^3 \over s_{1}^2t_{1}^2}\right)\right].
\end{eqnarray}

The square of the amplitude for the subprocess $c(p_1)+g(p_2)\to
\Xi_{cc}(p_3)+\bar{c}(p_4)$ with $(cc)$-diquark pair in $(cc)_{\bf
6}[^1S_0]$ can be written as,
\begin{eqnarray}
|\overline M|^2&=&{2^9{\alpha_s}^3|\Psi_{cc}(0)|^2{\pi}^4 \over
3^5{M}}\left[ M^2\left({-20 \over u_{1}^2}+{-1 \over
s_{1}t_{1}}+{-12u_{1}^2 \over s_{1}^2t_{1}^2}\right)+4M^4
\left({-12 \over s_{1}t_{1}u_{1}}+{-u_{1} \over s_{1}^2
t_{1}^2}+{-2u_{1}^3 \over s_{1}^3t_{1}^3}\right) \right.\nonumber  \\
&&\left.+ M^6\left({-48 \over s_{1}t_{1}u_{1}^2}+{8 \over
s_{1}^2t_{1}^2}+{-7u_{1}^2 \over s_{1}^3t_{1}^3}+{2u_{1}^4 \over
s_{1}^4t_{1}^4}\right)+2\left({-10 \over u_{1}}+{9u_{1} \over
s_{1}t_{1}}+{-2u_{1}^3 \over s_{1} ^2t_{1}^2}\right)\right].
\end{eqnarray}

The square of the amplitude for the subprocess $c(p_1)+c(p_2)\to
\Xi_{cc}(p_3) +g(p_4)$ with $(cc)$-diquark pair in $(cc)_{\bf\bar
3}[^3S_1]$ can be written as,
\begin{eqnarray}
|\overline M|^2 &=&{2^{11}{\alpha_s}^3|\Psi_{cc}(0)|^2{\pi}^4
\over 3^6{M}}\left[4M^2\left({-4s_{1}^4 \over t_{1}^3u_{1}^3}+
{11s_{1}^3 \over t_{1}^2u_{1}^3}+{15s_{1}^2 \over
t_{1}u_{1}^3}+{18s_{1} \over u_{1}^3}+{34t_{1}
 \over u_{1}^3}+{30t_{1}^2 \over s_{1}u_{1}^3}+{10
t_{1}^3 \over s_{1}^2u_{1}^3}\right) \right.\nonumber  \\
& &\left. +4M^4\left({20s_{1} ^3 \over t_{1}^3u_{1}^3}+{28s_{1}^2
\over t_{1}^2u_{1}^3}+{28s_{1} \over t_{1}u_{1}^3}+{17 \over
s_{1}t_{1} u_{1}}\right)+3M^6\left({-4s_{1}^4 \over
t_{1}^4u_{1}^4}+ {-14s_{1}^3 \over t_{1}^3u_{1}^4}+{-9s_{1}^2
\over t_{1}^2u_{1}^4}+{-2s_{1} \over
t_{1}u_{1}^4} \right.\right.\nonumber\\
& &\left.\left. +{-31 \over u_{1}^4} +{-36t_{1} \over
s_{1}u_{1}^4}+ {-12t_{1}^2 \over
s_{1}^2u_{1}^4}\right)+8\left({9s_{1}^3 \over
t_{1}^2u_{1}^2}+{11s_{1}^2 \over t_{1}u_{1}^2}+ {11s_{1} \over
u_{1}^2}+{2 \over s_{1}}\right)\right].
\end{eqnarray}

The square of the amplitude for the subprocess $c(p_1)+c(p_2)\to
\Xi_{cc}(p_3) +g(p_4)$ with $(cc)$-diquark pair in $(cc)_{\bf
6}[^1S_0]$ can be written as,
\begin{eqnarray}
|\overline M|^2&=&{2^{11}{\alpha_s}^3|\Psi_{cc}(0)|^2{\pi}^4 \over
3^6{M}}\left[M^2\left({12s_{1}^2 \over t_{1}^2u_{1}^2}+ {-s_{1}
\over t_{1}u_{1}^2}+{-1 \over u_{1}^2}+{20 \over
s_{1}^2}\right)+4M^4\left({2s_{1}^3\over t_{1}^3u_{1}^3}+{-s_{1}^2
\over t_{1}^2u_{1}^3}+{-s_{1} \over t_{1}u_{1}^3}+ \right.\right.\nonumber  \\
&&\left.\left. {12 \over s_{1}t_{1} u_{1}}\right)
+M^6\left({-2s_{1}^4 \over t_{1}^4u_{1}^4 }+{-7s_{1}^3 \over
t_{1}^3u_{1}^4}+{-15s_{1}^2 \over t_{1}^2u_{1}^4}+{-64s_{1} \over
t_{1}u_{1}^4}+ {-152 \over u_{1}^4}+{-144t_{1} \over s_{1}u_{1}^4
}+{-48t_{1}^2 \over
s_{1}^2u_{1}^4}\right)+ \right. \nonumber\\
&& \left. 2\left({2s_{1}^3 \over t_{1}^2u_{1}^2}+{9s_{1}^2 \over
t_{1}u_{1}^2}+{9s_{1} \over u_{1}^2}+{10 \over
s_{1}}\right)\right].
\end{eqnarray}
In these equation, $M$ is the mass of $\Xi_{cc}$ and $\Psi_{cc}(0)$
is the wavefunction at origin for the $[^3S_1]$ $cc$ state. And here
we have adopted that $h_3=|\Psi_{cc}(0)|^2$ and $h_1=h_3$.

\end{document}